\documentclass[bj]{imsart}

\RequirePackage[OT1]{fontenc}
\RequirePackage{amsthm,amsmath,natbib}
\RequirePackage[colorlinks,citecolor=blue,urlcolor=blue]{hyperref}

\usepackage{amsmath,epsfig}
\usepackage{latexsym}
\usepackage{amsfonts}
\usepackage{graphicx} 
\usepackage{subfigure}
\usepackage{amssymb}
\usepackage{stmaryrd}
\usepackage{caption}
\usepackage{color}
\usepackage{natbib}

\arxiv{1308.1883}

\startlocaldefs
\numberwithin{equation}{section}
\theoremstyle{plain}

\theoremstyle{remark}

\endlocaldefs

\newcommand{\dfn}{\triangleq}
\newcommand{\QED}{\Box} 
\newcommand{\rw}{\rightarrow} 
\newcommand{\Prob}{\mathbb{P}}    
\newcommand{\Real}{\mathbb{R}}  
  
\newcommand{\mB}{{\mathcal B}}
\newcommand{\mP}{{\mathcal P}}

\newcommand{\mI}{{\mathcal I}}

\newcommand{\mG}{{\mathcal G}}
\newcommand{\mF}{{\mathcal F}}

\newcommand{\mN}{{\mathcal N}}
\newcommand{\mU}{{\mathcal U}}
\newcommand{\mTN}{{\sf TN}}

\newcommand{\sff}{{\sf f}}
\newcommand{\intstep}{{\sf T}_e}

\newtheorem{Teorema}{\em Theorem}
\newtheorem{Definicion}{\em Definition}
\newtheorem{Lema}{\em Lemma}
\newtheorem{Proposicion}{\em Proposition} 

\newtheorem{Corolario}{\em Corollary}
\newtheorem{Assumption}{\em A.}
\newtheorem{Nota}{\em Remark}
\newtheorem{Algoritmo}{\em Algorithm}

\newtheorem{Result}{\em Result}

\begin{document}

\begin{frontmatter}
\title{Nested particle filters for online parameter estimation in discrete--time state--space Markov models}
\runtitle{Nested particle filters for online parameter estimation}

\begin{aug}
\author{\fnms{Dan} \snm{Crisan}\thanksref{a,e1}
\ead[label=e1,mark]{d.crisan@imperial.ac.uk}}
\and
\author{\fnms{Joaqu\'in} \snm{M\'iguez}\thanksref{b,e2}
\ead[label=e2,mark]{joaquin.miguez@uc3m.es}}

\address[a]{Department of Mathematics (Huxley Building), Imperial College London. 180 Queens Gate, London SW7 2BZ, UK.
\printead{e1}}

\address[b]{Department of Signal Theory and Communications, Universidad Carlos III de Madrid. Avenida de la Universidad 30, 28911 Legan\'es, Madrid, Spain.
\printead{e2}}


\runauthor{D. Crisan and J. M\'iguez}


\end{aug}

\begin{abstract}
We address the problem of approximating the posterior probability distribution of the fixed parameters of a state-space dynamical system using a sequential Monte Carlo method. The proposed approach relies on a nested structure that employs two layers of particle filters to approximate the posterior probability measure of the static parameters and the dynamic state variables of the system of interest, in a vein similar to the recent ``sequential Monte Carlo square'' (SMC$^2$) algorithm. However, unlike the SMC$^2$ scheme, the proposed technique operates in a purely recursive manner. In particular, the computational complexity of the recursive steps of the method introduced herein is constant over time. We analyse the approximation of integrals of real bounded functions with respect to the posterior distribution of the system parameters computed via the proposed scheme. As a result, we prove, under regularity assumptions, that the approximation errors vanish asymptotically in $L_p$ ($p \ge 1$) with convergence rate proportional to $\frac{1}{\sqrt{N}} + \frac{1}{\sqrt{M}}$, where $N$ is the number of Monte Carlo samples in the parameter space and $N\times M$ is the number of samples in the state space. This result also holds for the approximation of the joint posterior distribution of the parameters and the state variables. We discuss the relationship between the SMC$^2$ algorithm and the new recursive method and present a simple example in order to illustrate some of the theoretical findings with computer simulations.
\end{abstract}


\end{frontmatter}

\section{Introduction}




%
\subsection{Problem statement}

The problem of parameter estimation in state-space dynamical systems has received considerable attention, from different viewpoints \citep{Kitagawa98,Liu01b,Andrieu04,Kantas15,Carvalho10}, as it is almost ubiquitous in practical applications. In this paper, we investigate the use of particle filtering methods for the online Bayesian estimation of the static parameters of a state-space system. 

In order to ease the discussion, let us consider two (possibly vector-valued) random sequences $\{ X_t \}_{t=0,1,...}$ and $\{ Y_t \}_{t=1,2,...}$ representing the (hidden) state of a dynamical system and some related observations, respectively, with $t$ denoting discrete time. We assume that the state process is Markov and the observation $Y_t$ is independent of any other observations $\{ Y_k; k \ne t \}$, conditional on the state $X_t$. Both the conditional probability distribution of $X_t$ given the value of the previous state, $X_{t-1}=x_{t-1}$, and the probability density function (pdf) of $Y_t$ given $X_t=x_t$ are assumed to be known up to a vector of static (random) parameters, denoted by $\Theta$. These assumptions are commonly made in the literature and actually hold for many practical systems of interest (see, e.g., \citep{Ristic04,Cappe07}). Given a sequence of actual observations, $Y_1 = y_1, \ldots, Y_t = y_t, \ldots$, the goal is to track the posterior probability distributions of the state $X_t$, $t\ge 0$, and the parameter vector $\Theta$ over time.

In the sequel, we briefly review various existing approaches to the parameter estimation problem that involve particle filtering in some relevant manner. See \citep{Kantas15} for a more detailed survey of the field.

%
\subsection{Particle filters and parameter estimation}

When the parameter vector is given, i.e., $\Theta=\theta$ is known, the problem reduces to the standard stochastic filtering setting, which consists in tracking the posterior probability distribution of the state $X_t$, given the record of observations up to time $t>0$. In a few special cases (e.g., if the system is linear and Gaussian or the state-space is discrete and finite) there exist closed form solutions for the probability distribution of $X_t$ given $Y_1=y_1, ..., Y_t=y_t$, which is often termed the {\em filtering} distribution. However, analytical solutions do not exist for general, possibly nonlinear and non-Gaussian, systems and numerical approximation methods are then needed. One popular class of such methods are the so-called particle filters \citep{Gordon93,Kitagawa96,Liu98,Doucet00}. This is a family of recursive Monte Carlo algorithms that generate discrete random approximations of the sequence of probability measures associated to the filtering distributions at discrete time $t\ge 0$. 
   
Particle filters are well suited for solving the standard stochastic filtering problem. However,  the design of particle filters that can account for a random vector of parameters in the dynamic system (i.e., a static but unknown $\Theta$) has been an open issue for the past two decades. 

When the system of interest is endowed with some structure, there are some elegant techniques to handle the unknown parameters efficiently. For example, there are various conditionally-linear and Gaussian models that admit the analytical integration of $\Theta$ using the Kalman filter as an auxiliary tool, see, e.g., \citep{Doucet00,Chen00}. A similar approach can be taken with some non-Gaussian models appearing, e.g., in signal processing \citep{Bruno13}. In other cases, the analytical integration may not be feasible but the structure of the model can be such that the conditional probability law of $\Theta$ given $X_0=x_0, \ldots, X_t=x_t$ and $Y_1=y_1, \ldots, Y_t=y_t$ is tractable. In particular, if $\Theta$ depends on $X_{1:t} = \{ X_1, ..., X_t \}$ through a low-dimensional sufficient statistic then it is possible to draw efficiently from the posterior distribution of $\Theta$ (given $X_{0:t}=x_{0:t}$ and $Y_{1:t}=y_{1:t}$) \citep{Storvik02,Carvalho10} and then integrate the parameters out numerically.  

For arbitrary systems, with no particular structure, the more straightforward approach is to augment the state-space by including $\Theta$ as a constant-in-time state variable. This has been proposed in a number of forms and in various applications\footnote{It has also been proposed to use Markov chain Monte Carlo (MCMC) steps to prevent the collapse of the population representing the parameter posterior, that otherwise occurs due to the resampling steps \citep{Gilks01,Fearnhead02}.} but it can be shown that standard particle filters working on this augmented state-space do not necessarily converge in general because the resulting systems are non-ergodic \citep{Andrieu04,Papavasiliou06}. 
Another popular technique to handle static parameters within particle filtering consists in building a suitable kernel estimator of the posterior probability density function (pdf) of $\Theta$ given $Y_{1:t} = y_{1:t}$ from where new samples in the parameter space can be drawn \citep{Liu01b}. The latter step is often called ``rejuvenation'' or ``jittering'' (we adopt the latter term in the sequel). One key feature of this technique is the ``shrinkage'' of the density estimator in order to control the variance of the jittered particles. This method has been shown to work in some examples with low-dimensional $\Theta$, but has also been found to deliver poor performance in other simple setups \citep{Miguez05e}. A rigorous analysis of this technique is missing as well. 

Finally, there exists a large body of research on maximum likelihood estimation (MLE) for unknown parameters. Instead of handling $\Theta$ as a random variable and building an approximation of its posterior distribution, MLE techniques aim at computing a single-point estimate of the parameters. This is typically done by way of gradient optimisation methods, that lend themselves naturally to online implementations. A popular example is the recursive maximum likelihood (RML) algorithm \citep{LeGland97,Poyiadjis11,DelMoral15}. As an alternative to gradient search methods, expectation maximization (EM) techniques have also been proposed for the optimisation of the parameter likelihood, both in offline and online versions \citep{Andrieu04,Kantas15}. These techniques use particle filtering as an ancillary tool to approximate either the gradient of the likelihood function \citep{DelMoral15} or some sufficient statistics \citep{Andrieu04} and have been advocated as more robust than those based on state-space augmentation, artificial evolution or kernel density estimation \citep{Andrieu04,Kantas15}. 

%
\subsection{Non-recursive methods}

A number of new methods related to particle filtering have been proposed in the past few years that tackle the problem of approximating the distribution of the parameter vector $\Theta$ given the observations $Y_{1:T}=y_{1:T}$. These techniques include the iterated batch importance sampling (IBIS) algorithm of \citep{Chopin02} and extensions of it that rely on the {\em nesting} of particle methods (such as in \citep{Papavasiliou06} or \citep{Chopin12}), combinations of Markov chain Monte Carlo (MCMC) and particle filtering \citep{Andrieu10}, variations of the population Monte Carlo methodology \citep{Koblents13} and particle methods for the approximation of the parameter likelihood function \citep{Olsson08}.

The IBIS method is a sequential Monte Carlo (SMC) algorithm that updates a population of samples $\theta_t^{(i)}$, $i=1, ..., N$, in the space of $\Theta$, with associated importance weights, at every time step. The technique involves regular MCMC steps, in order to rejuvenate the population of samples, and the ability to compute the pdf of every observation variable $Y_t$, given the previous observation record $Y_{1:t-1}=y_{1:t-1}$ and a fixed value of the parameters, $\Theta = \theta$. Let us denote such densities as $d(y_t|y_{1:t-1},\theta)$ for the sake of conciseness. The need to obtain $d(y_t|y_{1:t-1},\theta)$ in closed-form has two important implications. First, IBIS is not a recursive algorithm, since each  time we need to compute $d(y_t|y_{1:t-1},\theta)$ for a new sample point $\Theta=\theta$ in the parameter space it is necessary to process the entire sequence of observations $Y_{1:t-1}=y_{1:t-1}$. Second, the algorithm can only be applied when the dynamic model has some suitable structure (e.g., the system may be linear and Gaussian conditional on $\Theta$) that enables us to actually find $d(y_t|y_{1:t-1},\theta)$ in closed form.    

In \citep{Papavasiliou06}, these difficulties with the IBIS method are addressed by using two layers of Monte Carlo methods. First, a random grid of points in the space of $\Theta$, say $\theta^{(1)}, ..., \theta^{(N)}$, is generated. Then, for each $\Theta=\theta^{(i)}$, $i=1, ..., N$, a particle filter is employed targeting the signal $\{ X_t \}_{t=0, 1, ...}$. The latter particle filters provide approximations of $d(y_t|y_{1:t-1},\theta^{(i)})$, $i=1, ..., N$, and, since the grid in the parameter space is fixed, a single sweep over the observations $Y_{1:T}=y_{1:T}$, $T<\infty$, is sufficient, hence the algorithm is recursive. The practical weakness of this approach is that the random grid over the parameter space is generated a priori (irrespective of the observations $Y_{1:T}=y_{1:T}$) and it is not updated as the observations are processed. Therefore, when the prior distribution of $\Theta$ differs from the posterior distribution (of $\Theta$ conditional on $Y_{1:T}=y_{1:T}$) significantly, a very large number, $N$, of samples in the parameter space is needed to guarantee a fair performance. 

The methodology proposed in \citep{Chopin12} is also an extension of the IBIS technique. Similarly to the method in \citep{Papavasiliou06}, a random grid is created over the parameter space and a particle filter is run for every node in the grid. However, unlike the technique in \citep{Papavasiliou06}, the grid of samples in the space of $\Theta$ is updated over time, as the batch of observations $Y_{1:T}=y_{1:T}$ is processed. In particular, if $\{ \theta_{t-1}^{(i)}, i=1, ..., N\}$ is the grid at time $t-1$, a particle filter is used to process $y_t$ and then a new grid $\{ \theta_t^{(i)}, i=1, ..., N\}$ can be generated. This filter involves the computation of weights that depend on the densities $d(y_t|y_{1:t-1},\theta_t^{(i)})$, $i=1, ..., N$ (similar to the original IBIS). For each point $\Theta=\theta_t^{(i)}$, a particle filter is run to approximate $d(y_t|y_{1:t-1},\theta_t^{(i)})$. The resulting method is called SMC$^2$ in \citep{Chopin12} because of the two nested layers of particle filters. It is more flexible and general than the original IBIS and its extension in \citep{Papavasiliou06}, but it is not a recursive algorithm. New samples in the parameter space are generated by way of particle MCMC \citep{Andrieu10} (see below) moves and resampling steps in order to avoid the degeneracy of the particle filter. However, each time a new point in the parameter space is generated at time $t$, say $\theta_t'$, a new filter has to be run from time 0 to time $t$. Therefore, the computationally complexity of the method grows quadratically with time. A major advantage of the SMC$^2$ algorithm is that the approximation errors vanish asymptotically as the number of samples $N$ on the parameter space increases, independently of the number of particles used to approximate the densities $d(y_t|y_{1:t-1},\theta_t^{(i)})$ in the second layer of particle filters, which can stay fixed. This is shown in \citep{Chopin12} resorting to a well known unbiasedness property proved in \citep{DelMoral04}. 


A technique that has quickly gained popularity for parameter estimation is the particle MCMC method of \citep{Andrieu10} (employed as a building block for the SMC$^2$ method described above). It essentially consists in an MCMC algorithm to approximate the posterior distribution of $\Theta$ given $Y_{1:t}=y_{1:t}$. Such construction is intractable if addressed directly because the likelihoods $d(y_{1:t}|\theta)$ cannot be conmputed exactly. To circumvent this difficulty, it was proposed in \citep{Andrieu10} to use particle filters in order to approximate them. The same trick has been used in the population Monte Carlo \citep{Cappe04} framework to tackle the approximation of the posterior distribution of $\Theta$ using particles with nonlinearly transformed weights \citep{Koblents13}. The latter technique has been reported to be computationally more efficient than particle MCMC methods in some examples. These two types of algorithms, as well as the SMC$^2$ scheme, revolve around the ability to approximate the factors $d(y_t|y_{1:t-1},\theta)$ using particle filtering. 

An alternative, and conceptually simple, approach to compute the likelihood of $\Theta$ given $Y_{1:t}$ has been proposed in \citep{Olsson08}. The problem is addressed by generating a random grid over the parameter space (either random or deterministic, but fixed), then using particle filters to compute the value of the likelihood at each node and finally obtaining an approximation of the whole function by interpolating the nodes. If a point estimate of the parameters is needed, standard optimisation techniques can be applied to the interpolated approximation. Convergence of the $L_p$ error norms is proved in \citep{Olsson08} for problems where both the parameter space and the state space are compact.

%
\subsection{Contributions}

We introduce a particle filtering method for the approximation of the joint posterior distribution of the signal and the unknown parameters, $X_t$ and $\Theta$, respectively, given the data $Y_{1:t}=y_{1:t}$. Similar to \citep{Papavasiliou06} and \citep{Chopin12}, the algorithm consists of two nested layers of particle filters: an ``outer'' filter that approximates the probability measure of $\Theta$ given the observations and a set of ``inner'' filters, one per sample generated in the outer filter, that yield approximations of the posterior measures that result for $X_t$ conditional on the observations {\em and} each specific sample of $\Theta$. The outer filter directly provides an approximation of the marginal posterior distribution of $\Theta$, whereas a suitable combination of the latter with the outcomes of the inner filters yields an approximation of the joint posterior probability measure of $X_t$ and $\Theta$.

The method is very similar to the SMC$^2$ scheme of \citep{Chopin12} in its structure. However, unlike SMC$^2$, it is a purely recursive procedure and, therefore, it is more suitable for an online implementation. At every time step, all the probability measure approximations (both marginal and joint) are updated recursively, with a fixed computational cost. Also, the jittering of particles in the SMC$^2$ algorithm is carried out using a particle MCMC kernel \citep{Chopin12}, that leaves the target distribution invariant but cannot be implemented recursively, while the proposed scheme works with simpler Markov kernels easily amenable to online implementations. A detailed comparison between the proposed algorithm and the SMC$^2$ method of \citep{Chopin12} is presented in Section \ref{ssComparison}.

The core of the paper is devoted to the analysis of the proposed algorithm. We study the approximation, via the nested particle filtering scheme, of 1-dimensional statistics of the posterior distribution of the system parameters. Under regularity assumptions, we prove that the $L_p$ norms of the approximation errors vanish with rate proportional to $\frac{1}{\sqrt{N}} + \frac{1}{\sqrt{M}}$, where $N$ and $N \times M$ are the number of samples in the parameter space and the number of particles in the state space, respectively. This result also holds for the approximation of the joint posterior distribution of the parameters and the state variables. 

The analysis builds upon two basic assumptions, which determine the applicability of the algorithm. The most important one is that the optimal filter for the state space model of interest is continuous with respect to (w.r.t.) the parameter $\theta$, i.e., that small changes to the parameter lead to small changes to the posterior probability measure of the state given the available observations. It is this continuity property that makes the implementation of the proposed recursive algorithm feasible and determines some key practical elements of the algorithm, including the magnitude of the jittering of the particles. Non-recursive methods, such as particle MCMC or SMC$^2$, are not subject to this constraint. The second basic assumption is that the parameter space is a compact set and the the conditional pdf of the observations is well behaved (positive and upper bounded) uniformly over that set. The proposed technique is not guaranteed to work if the parameters have to be searched over an infinite support or, most importantly, if the conditional pdf of the observations has some singularity (e.g., it becomes unbounded) for some parameter values.

To complement the analysis, we also provide a numerical example, where we apply the proposed algorithm to jointly track the state variables and estimate the fixed parameters of a (stochastic version of the) Lorenz 63 system. The length of the observation periods for this example ($\sim 40,000$ discrete time steps) is large enough to make the application of the non-recursive SMC$^2$ method impractical, while the proposed technique attains accurate estimates of the unknown parameters and tracks the state variables closely.

\subsection{Organisation of the paper}
 
We present a general description of the random state-space Markov models of interest in this paper in Section \ref{sModel}, including a brief review of the standard particle filter with known parameters. The recursive nested particle filter scheme is introduced in Section \ref{sNested}. In Section \ref{sResults} we provide a summary of the main theoretical properties of the proposed algorithm and discuss how it compares to the (non recursive) SMC$^2$ method of \citep{Chopin12}. The analysis of the approximation errors in $L_p$ is contained in Section \ref{sConvergence}, together with a brief discussion on the computation of an effective sample size for the proposed algorithm. Section \ref{sExamples} presents some illustrative numerical results for a simple example and, finally, Section \ref{sConclusions} is devoted to the conclusions.

\section{Background} \label{sModel}

\subsection{Notation and preliminaries}

We first introduce some common notation to be used through the paper, broadly classified by topics. Below, $\Real$ denotes the real line, while for an integer $d\ge 1$, $\Real^d=\overbrace{\Real \times \ldots \times \Real}^{d \mbox{ {\tiny times}}}$.

\begin{itemize}

\item Functions. Let $S \subseteq \Real^d$ be a subset of $\Real^d$.
        \begin{itemize}
        \item The supremum norm of a real function $f:S \rw \Real$ is denoted as $\| f \|_\infty = \sup_{x\in S} | f(x) |$.
        \item $B(S)$ is the set of bounded real functions over $S$, i.e., $f \in B(S)$ if, and only if, $\| f \|_\infty < \infty$.
        \end{itemize}
        
\item Measures and integrals. 
        \begin{itemize}
        \item $\mB(S)$ is the $\sigma$-algebra of Borel subsets of $S$.
        \item $\mP(S)$ is the set of probability measures over the measurable space $(\mB(S),S)$.
        \item $(f,\mu) \dfn \int f(x) \mu(dx)$ is the integral of a real function $f:S \rw \Real$ w.r.t. a measure $\mu \in \mP(S)$.
        \item Given a probability measure $\mu \in \mP(S)$, a Borel set $A \in \mB(S)$ and the indicator function 
        $$
        I_A(x) = \left\{
                \begin{array}{ll}
                1, &\mbox{if } x \in A\\
                0, &\mbox{otherwise}
                \end{array}
        \right.,
        $$
        $\mu(A) = (I_A,\mu) = \int I_A(x) \mu(dx)$ 
        is the probability of $A$.

        \end{itemize}

        
\item Sequences, vectors and random variables (r.v.).
        \begin{itemize}
        \item We use a subscript notation for finite sequences, namely $x_{t_1:t_2} \dfn \{ x_{t_1}, \ldots, x_{t_2} \}$.
        \item For an element $x=(x_1,\ldots,x_d) \in \Real^d$ of an Euclidean space, its norm is denoted as $\| x \| = \sqrt{ x_1^2+\ldots+x_d^2 }$.
        \item Let $Z$ be a r.v. taking values on $\Real^d$, with associated probability measure $P \in \mP(\Real^d)$. The $L_p$ norm of $Z$, with $p \ge 1$, is $\| Z \|_p \dfn E[ |Z|^p ]^{1/p} = \left( \int |z|^p P(dz) \right)^{\frac{1}{p}}$ (where $E[\cdot]$ denotes expectation).
        \end{itemize}
\end{itemize}

\begin{Nota}
Let $\alpha, \beta, \bar \alpha, \bar \beta \in \mP(S)$ be probability measures and let $f,h \in B(S)$ be two real bounded functions on $S$ such that $(h,\bar \alpha)>0$ and $(h,\bar \beta)>0$. If the identities 
$$
(f,\alpha) = \frac{
	(fh,\bar \alpha)
}{
	(h,\bar \alpha)
} \quad \mbox{and} \quad (f,\beta) = \frac{
	(fh,\bar \beta)
}{
	(h,\bar \beta)
}
$$
hold, then it is straightforward to show (see, e.g., \citep{Crisan01}) that
\begin{equation}
| (f,\alpha)-(f,\beta) | \le \frac{
	1
}{
	(h,\bar \alpha)
} \left|
	(fh,\bar \alpha) - (fh,\bar \beta)
\right| + \frac{
	\| f \|_\infty
}{
	(h,\bar \alpha)
} \left|
	(h,\bar \alpha) - (h,\bar \beta)
\right|. 
\label{eqPreliminaries}
\end{equation}
\end{Nota}

\subsection{State-space Markov models in discrete time} \label{ssModel}

Consider two random sequences, $\{ X_t \}_{t \ge 0}$ and $\{ Y_t \}_{t \ge 1}$ taking values in $\Real^{d_x}$ and $\Real^{d_y}$, respectively, and a r.v.  $\Theta$ taking values on a compact set $D_\theta \subset \Real^{d_\theta}$. Let $\Prob_t$ be the joint probability measure for the triple $\left( \{X_k\}_{0 \le k \le t}, \{Y_k\}_{0 < k < t}, \Theta \right)$, that we assume to be absolutely continuous w.r.t. the Lebesgue measure on $\mB(\Real^{d_x(t+1)} \times \Real^{d_y t} \times D_\theta)$.

We refer to the  sequence $\{ X_t \}_{t\ge 0}$ as the state (or signal) process and we assume that it is an inhomogeneous Markov chain governed by an initial probability measure $\tau_{0} \in \mP(\Real^{d_x})$ and a sequence of transition kernels $\tau_{t,\theta} : \mB(\Real^{d_x}) \times \Real^{d_x} \rw [0,1]$ indexed by a realisation of the r.v. $\Theta=\theta$. To be specific, we define
\begin{eqnarray}
\tau_{0}(A) &\dfn& \Prob_0\left\{ X_0 \in A \right\}, \label{eqPrior} \\ 
\tau_{t,\theta}(A|x_{t-1}) &\dfn& \Prob_t\left\{ X_t \in A | X_{t-1}=x_{t-1}, \Theta=\theta \right\}, \quad t \ge 1, \label{eqKernel}
\end{eqnarray}
where $A \in \mB(\Real^{d_x})$ is a Borel set. The sequence $\{ Y_t \}_{t \ge 1}$ is termed the observation process. Each r.v. $Y_t$ is assumed to be conditionally independent of other observations given $X_t$ and $\Theta$, namely
\begin{equation}
\Prob_t\left\{ Y_t \in A | X_{0:t} = x_{0:t}, \Theta = \theta, \{ Y_k = y_k \}_{k \ne t} \right\} = \Prob_t\left\{ Y_t \in A | X_t = x_t, \Theta=\theta  \right\} \nonumber
\end{equation}
for any $A \in \mB(\Real^{d_y})$. Additionally, we assume that every probability measure $\gamma_{t,\theta} \in \mP(\Real^{d_y})$ in the family 
\begin{equation}
\gamma_{t,\theta}(A|x_t) \dfn \Prob_t\left\{ Y_t \in A | X_t=x_t, \Theta=\theta \right\}, \quad A \in \mB(\Real^{d_x}), \quad \theta \in D_\theta, \quad t \ge 1,
\label{eqObservations1}
\end{equation}
has a nonnegative density w.r.t. the Lebesgue measure. The function $g_{t,\theta}(y|x) \ge 0$ is proportional to this density, hence we write 
\begin{equation}
\gamma_{t,\theta}(A|x_t) = \int c I_A(y) g_{t,\theta}(y|x_t)dy,
\label{eqObservations2}
\end{equation}
where $c$ is a (possibly unknown) normalisation constant, assumed independent of $y$, $x$ and $\theta$. 

The prior $\tau_0$, the kernels $\{ \tau_{t,\theta} \}_{t \ge 1}$, and the functions $\{ g_{t,\theta} \}_{t \ge 1}$, describe a stochastic Markov state-space model in discrete time. Note that the model is indexed by $\theta \in D_\theta$, which is henceforth termed the system parameter. The a priori probability measure of the r.v. $\Theta$ is denoted $\mu_0$, i.e., for any $A \in \mB(D_\theta)$, $\mu_0(A) \dfn \Prob_0\{ \theta \in A \}$. 

If $\Theta=\theta$ (the parameter is given), then the stochastic filtering problem consists in the computation of the posterior probability measure of the state $X_t$ given the parameter and a sequence of observations up to time $t$. Specifically, for a given observation record $\{ y_t \}_{t \ge 1}$, we seek the measures
\begin{equation}
\phi_{t,\theta}(A) \dfn \Prob_t\left\{ X_t \in A | Y_{1:t}=y_{1:t}, \Theta=\theta \right\}, \quad t=0, 1, 2, ...
\nonumber
\end{equation}
where $A \in \mB(\Real^{d_x})$. For many practical problems, the interest actually lies in the computation of statistics of the form $(f,\phi_{t,\theta})$ for some integrable function $f:\Real^{d_x}\rw\Real$. Note that, for $t=0$, we recover the prior signal measure, i.e., $\phi_{0,\theta}=\tau_{0}$ independently of $\theta$.

There are many applications in which the parameter $\Theta$ is unknown and the goal is to fit the model using a given sequence of observations. In that case, the sequence of probability measures of interest is
$$
\mu_t(A) \dfn \Prob_t\left\{ \Theta \in A | Y_{1:t} = y_{1:t} \right\}, \quad t = 0, 1, 2, ..., \mbox{ where } A \in \mB(D_\theta).
$$
If both the fitting of the model and the tracking of the state variables $\{ X_t \}_{t\ge 0}$ are sought, then we need to approximate the joint probability measures
$$
\pi_t(A \times A') \dfn \Prob_t\left\{ X_t \in A, \Theta \in A' | Y_{1:t}=y_{1:t} \right\}, \quad t = 0, 1, 2, ...,
$$ 
where $A \in \mB(\Real^{d_x})$ and $A' \in \mB(D_\theta)$. Note that we can write the joint measure $\pi_t$ as a function of the marginals $\phi_{t,\theta}$ and $\mu_t$. Indeed, if given $A \in \mB(\Real^{d_x})$ we introduce the real function $\sff_t^A : D_\theta \rw [0,1]$, where $\sff_t^A(\theta) = \phi_{t,\theta}(A)$, then
\begin{equation}
\pi_t(A \times A') = ( I_{A'} \sff_t^A, \mu_t ) 
= \int I_{A'}(\theta) \sff_t^A(\theta) \mu_t(d\theta)
= \int \int I_{A'}(\theta) I_{A}(x) \phi_{t,\theta}(dx) \mu_t(d\theta). \label{eqJointDef}
\end{equation}

\subsection{Standard particle filter} \label{ssPFs}

Assume that both the parameter $\Theta=\theta$ and a sequence of observations $Y_{1:T} = y_{1:T}$, $T<\infty$, are fixed. Then, the sequence of measures $\{ \phi_{t,\theta} \}_{t \ge 1}$ can be numerically approximated using particle filtering. Particle filters are numerical methods based on the recursive relationship between $\phi_{t,\theta}$ and $\phi_{t-1,\theta}$. In particular, let us introduce the predictive measure $\xi_{t,\theta} \dfn \tau_{t,\theta}\phi_{t-1,\theta}$ such that, for any integrable function $f:\Real^{d_x}\rw\Real$, we obtain
\begin{equation}
(f,\xi_{t,\theta}) = \int \int f(x)\tau_{t,\theta}(dx|x')\phi_{t-1,\theta}(dx') 
=  \left(
	(f,\tau_{t,\theta}), \phi_{t-1,\theta}
\right),
\label{eqPredicting}
\end{equation}
where we note that $\int f(x)\tau_{t,\theta}(dx|x')$ is itself a map $\Real^{d_x}\rw\Real$. Integrals w.r.t. the filter measure $\phi_{t,\theta}$ can be rewritten by way of $\xi_{t,\theta}$ as 
\begin{equation}
(f,\phi_{t,\theta}) = \frac{
	(fg_{t,\theta}^{y_t},\xi_{t,\theta})
}{
	(g_{t,\theta}^{y_t},\xi_{t,\theta})
},
\label{eqAlternative}
\end{equation}
where $g_{t,\theta}^{y_t}(x) \dfn g_{t,\theta}(y_t|x)$ is the likelihood of $x \in \Real^{d_x}$. Eqs. \eqref{eqPredicting} and \eqref{eqAlternative} are used extensively through the paper. They are instances of the Chapman-Kolmogorov equation and the Bayes theorem, respectively.

The simplest particle filter, often called `standard particle filter' or `bootstrap filter' \citep{Gordon93} (see also \citep{Doucet01}), can be described as follows.

\begin{Algoritmo} \label{alBF}
Bootstrap filter conditional on $\Theta=\theta$.
\begin{enumerate}
\item {\sf Initialisation.} At time $t=0$, draw $N$ i.i.d. samples,  $x_0^{(i)}$, $n=1,\ldots,N$, from the prior $\tau_0$.
 
\item {\sf Recursive step.} Let $\{ x_{t-1}^{(n)} \}_{1 \le n \le N}$ be the particles (Monte Carlo samples) generated at time $t-1$. At time $t$, proceed with the two steps below.
        \begin{enumerate}
        \item For $n=1,...,N$, draw a sample $\bar x_t^{(n)}$ from the probability distribution $\tau_{t,\theta}(\cdot|x_{t-1}^{(n)})$ and compute the normalised weight
        \begin{equation}
        w_t^{(n)} = \frac{
                g_{t,\theta}^{y_{t}}(\bar x_t^{(n)})
        }{
                \sum_{k=1}^N g_{t,\theta}^{y_{t}}(\bar x_t^{(k)})
        }.
        \end{equation}
        
        \item For $n=1,...,N$, let $x_t^{(n)}=\bar x_t^{(k)}$ with probability $w_t^{(k)}$, $k \in \{1,...,N\}$.
        \end{enumerate}
\end{enumerate}
\end{Algoritmo}

Step 2.(b) is referred to as resampling or selection. In the form stated here, it reduces to the so-called multinomial resampling algorithm \citep{Doucet00,Douc05} but  convergence of the filter can be easily proved for various other schemes (see, e.g., the treatment of the resampling step in \citep{Crisan01}). Using the set $\{ x_t^{(n)} \}_{1 \le n \le N}$, we construct random approximations of $\xi_{t,\theta}$ and $\phi_{t,\theta}$, namely
\begin{equation}
\xi_{t,\theta}^N(dx_t) = \frac{1}{N} \sum_{n=1}^N \delta_{\bar x_t^{(n)}}(dx_t) \quad \mbox{and} \quad
\phi_{t,\theta}^N(dx_t) = \frac{1}{N} \sum_{n=1}^N \delta_{x_t^{(n)}}(dx_t),
\end{equation}
where $\delta_{x_t^{(n)}}$ is the Dirac delta measure located at $X_t=x_t^{(n)}$. For any integrable function $f$ in the state space, it is straightforward to approximate the integrals $(f,\xi_{t,\theta})$ and $(f,\phi_{t,\theta})$ as 
\begin{equation}
(f,\xi_{t,\theta}) \approx (f,\xi_{t,\theta}^N) = \frac{1}{N} \sum_{n=1}^N f(\bar x_t^{(n)}) \quad \mbox{and} \quad
(f,\phi_{t,\theta}) \approx (f,\phi_{t,\theta}^N) = \frac{1}{N} \sum_{n=1}^N f(x_t^{(n)}),
\label{eqApprox_of_f}
\end{equation} 
respectively.

The convergence of particle filters has been analysed in a number of different ways. Here we use results for the convergence of the $L_p$ norms ($p\ge 1$) of the approximation errors. 

\begin{Teorema} \label{thConvPF-0}
Assume that both the system parameter $\Theta=\theta$ and the sequence of observations $Y_{1:T}=y_{1:T}$ are fixed (with $T<\infty$), $g_{t,\theta}^{y_{t}} \in B(\Real^{d_x})$ and $g_{t,\theta}^{y_t}>0$ (in particular, $(g_{t,\theta}^{y_{t}},\xi_{t,\theta}) > 0$) for every $t=1,2,...,T$. Then for any $f \in B(\Real^{d_x})$, any $p \ge 1$ and every $t=1, \ldots, T$, 
\begin{eqnarray}
\left\|
        (f,\xi_{t,\theta}^N) - (f,\xi_{t,\theta})
\right\|_p 
\le
        \frac{
                \bar c_{t,\theta} \| f \|_\infty
        }{
                \sqrt{N}        
        } \quad \mbox{and} \quad
\left\|
        (f,\phi_{t,\theta}^N) - (f,\phi_{t,\theta})
\right\|_p 
\le 
        \frac{
                c_{t,\theta} \| f \|_\infty
        }{
                \sqrt{N}        
        }, \nonumber
\end{eqnarray}
where $\bar c_{t,\theta}, c_{t,\theta}<\infty$ are constants independent of $N$, $\| f \|_\infty = \sup_{x \in \Real^{d_x}} |f(x)|$ and the expectations are taken over the distributions of the random measures $\xi_{t,\theta}^N$ and $\phi_{t,\theta}^N$, respectively. 
\end{Teorema}

\noindent {\bf Proof:} This result is a special case of, e.g., Lemma 1 in \citep{Miguez13b}. $\QED$


Theorem \ref{thConvPF-0} is fairly standard. A similar proposition was already proved in \citep{DelMoral00}, albeit under additional assumptions on the state-space model, and bounds for $p=2$ and $p=4$ can also be found in a number of references (see, e.g., \citep{Crisan01,Crisan02,DelMoral04}). 
It is also possible to establish conditions that make the convergence result of Theorem \ref{thConvPF-0} uniform over the parameter space. Recall that the r.v. $\Theta$ has compact support $D_\theta \subset \Real^{d_\theta}$ and denote
\begin{eqnarray}
\| g_t^{y_t} \|_\infty &\dfn& \sup_{\theta \in D_\theta} \| g_{t,\theta}^{y_t} \|_\infty, \\
u_t(\theta) &\dfn& (g_{t,\theta}^{y_t}, \xi_{t,\theta}) \quad \mbox{and} \\
u_{t,\inf} &\dfn& \inf_{\theta \in D_\theta} u_t(\theta).
\end{eqnarray} 
We can state a result very similar to Theorem \ref{thConvPF-0}, but with the constant in the upper bound of the approximation error being independent of the parameter $\theta$. For convenience in the exposition of the rest of the paper, we first establish the convergence, uniform over the parameter space $D_\theta$, of the recursive step in the particle filter.

\begin{Lema} \label{lm1StepUnifConv}
Choose any $\theta \in D_\theta$ and any $f \in B(\Real^{d_x})$. Assume that the sequence of observations $Y_{1:t}=y_{1:t}$ is fixed (for some $t<\infty$) and a discrete random measure
$
\phi_{t-1,\theta}^N(dx_{t-1}) = \frac{1}{N} \sum_{n=1}^N \delta_{x_{t-1}^{(n)}}(dx_{t-1})
$
is available such that, for any $p \ge 1$, 
\begin{equation}
\| (f,\phi_{t-1,\theta}^N) - (f,\phi_{t-1,\theta}) \|_p \le \frac{
	c_{t-1} \| f \|_\infty
}{
	\sqrt{N}
} + \frac{
	\bar c_{t-1} \| f \|_\infty
}{
	\sqrt{M}
} ,
\label{eqInductionHypo}
\end{equation}
where $M \ge 1$ is an integer and $c_{t-1}, \bar c_{t-1}<\infty$ are constants independent of $N$, $M$ and $\theta$.

If $g_{t,\theta}^{y_t}>0$, $\| g_{t}^{y_{t}} \| < \infty$ and $u_{t,\inf} > 0$, then, for any $p \ge 1$,
\begin{eqnarray}
\left\|
        (f,\xi_{t,\theta}^N) - (f,\xi_{t,\theta})
\right\|_p &\le& 
        \frac{
                \tilde c_{t} \| f \|_\infty
        }{
                \sqrt{N}        
        } +  \frac{
                \bar {\tilde c}_{t} \| f \|_\infty
        }{
                \sqrt{M}        
        }
\quad \mbox{and} \nonumber\\
\left\|
        (f,\phi_{t,\theta}^N) - (f,\phi_{t,\theta})
\right\|_p &\le& 
        \frac{
                c_{t}\| f \|_\infty
        }{
                \sqrt{N}        
        } + \frac{
                \bar c_{t}\| f \|_\infty
        }{
                \sqrt{M}        
        }, \nonumber
\end{eqnarray}
where $\xi_{t,\theta}^N$ and $\phi_{t,\theta}^N$ are computed as in the recursive step of the standard particle filter, $\tilde c_t$, $\bar {\tilde c}_t$, $c_{t}$ and $\bar c_t$ are finite constants independent of $N$, $M$ and $\theta$, and the expectations are taken over the distributions of the random measures $\xi_{t,\theta}^N$ and $\phi_{t,\theta}^N$. If $\bar c_{t-1}=0$ then $\bar c_t=\bar {\tilde c}_t=0$. 
\end{Lema}

\noindent {\bf Proof:} See Appendix \ref{ap1StepUnifConv}. $\QED$

The (arbitrary) integer $M$ introduced for notational convenience and the error term $\propto \frac{1}{\sqrt{M}}$ plays no role in the proof of Lemma \ref{lmUnifConvPF} below. It is included exclusively to ease the exposition of some proofs in Section \ref{sConvergence}. Given Lemma \ref{lm1StepUnifConv}, it is straightforward to establish the convergence, uniform over $D_\theta$, of the standard particle filter.

\begin{Lema} \label{lmUnifConvPF}
Assume that the sequence of observations $Y_{1:T}=y_{1:T}$ is fixed (for some $T<\infty$), $g_{t,\theta}^{y_t}>0$, $\| g_{t}^{y_{t}} \| < \infty$ and $u_{t,\inf} > 0$ for every $t=1,2,...,T$. Then, for any $f \in B(\Real^{d_x})$, any $\theta \in D_\theta$ and any $p \ge 1$,
\begin{eqnarray}
\left\|
        (f,\xi_{t,\theta}^N) - (f,\xi_{t,\theta})
\right\|_p \le 
        \frac{
                \tilde c_{t} \| f \|_\infty
        }{
                \sqrt{N}        
        } \quad \mbox{and} \quad
\left\|
        (f,\phi_{t,\theta}^N) - (f,\phi_{t,\theta})
\right\|_p \le 
        \frac{
                c_{t}\| f \|_\infty
        }{
                \sqrt{N}        
        } \nonumber
\end{eqnarray}
for $t=0,1,\ldots, T$, where $\tilde c_t(f)$ and $c_{t}(f)$ are finite constants, independent of both $N$ and $\theta$, and the expectations are taken over the distributions of the random measures $\xi_{t,\theta}^N$ and $\phi_{t,\theta}^N$. 
\end{Lema}
 
\noindent {\bf Proof:} See Appendix \ref{apProof_of_Lemma_2}. $\QED$

\begin{Nota} \label{rmF_of_theta}
Lemmas \ref{lm1StepUnifConv} and \ref{lmUnifConvPF} also hold for any test function $f^\theta:\Real^{d_x}\rw\Real$ (i.e., dependent on $\theta$) as long as the upper bounds
$$
\| f \|_\infty  = \sup_{\theta \in D_\theta} \| f^\theta \|_\infty, 
\quad \mbox{and} \quad
\| g_t^{y_t} \|_\infty = \sup_{\theta \in D_\theta} \| g_{t,\theta}^{y_t} \|_\infty
$$
are finite and the lower bound
$
\inf_{\theta \in D_\theta} g_{t,\theta}^{y_t}(x)
$
is positive for every $x \in \Real^{d_x}$ and every $t = 1, ..., T$. Note that $
\inf_{\theta \in D_\theta} g_{t,\theta}^{y_t}(x) > 0
$ implies that $u_{t,inf} = \inf_{\theta \in D_\theta} u_t(\theta) > 0$. Under these assumptions the constants $c_t$ and $\bar c_t$ in the statement of Lemma \ref{lm1StepUnifConv} are independent of $\theta$ (they depend on $u_{t,\inf}$ and $\|g_t^{y_t}\|_\infty$, though).
\end{Nota}

\section{Nested particle filter} \label{sNested}


\subsection{Sequential importance sampling in the parameter space} 

We aim at devising a recursive algorithm that generates approximations of the posterior probability measures $\mu_t(d\theta)$, $t=1, 2, ...$, using a sequential importance sampling scheme. The key object needed to attain this goal is the marginal likelihood of the parameter $\Theta$ at time $t$, i.e., the conditional probability density of the observation $Y_t$ given a parameter value $\Theta=\theta$ and a record of observations $Y_{1:t-1}=y_{1:t-1}$.  
 
To be specific, assume that the observations $Y_{1:t-1}=y_{1:t-1}$ are fixed and let
\begin{equation}
\upsilon_{t,\theta}(A) \dfn \Prob_t\left\{ 
	Y_t \in A | Y_{1:t-1}=y_{1:t-1}, \Theta = \theta 
\right\}, \quad A \in \mB(\Real^{d_y}),
\nonumber
\end{equation}
be the probability measure associated to the (random) observation $Y_t$ conditional on $Y_{1:t-1}=y_{1:t-1}$ and the parameter vector $\Theta=\theta$. Let us assume that $\upsilon_{t,\theta}$ has a density $u_{t,\theta}:\Real^{d_y} \rw [0,+\infty)$ w.r.t. the Lebesgue measure, i.e.,
\begin{equation}
\upsilon_{t,\theta}(A) = \int I_A(y) u_{t,\theta}(y) dy, \quad \mbox{for any $A \in \mB(\Real^{d_y})$.}
\nonumber
\end{equation}
When the actual obsevation $Y_t=y_t$ is collected, the density $u_{t,\theta}(y_t)$ can be evaluated as an integral, namely 
$u_{t,\theta}(y_t) = (g_{t,\theta}^{y_t}, \xi_{t,\theta})$, and it yields the marginal likelihood of the parameter value $\theta$, denoted as
\begin{equation}
u_t(\theta) \dfn u_{t,\theta}(y_t) = (g_{t,\theta}^{y_t}, \xi_{t,\theta}).
\nonumber
\end{equation}

A straightforward Monte Carlo approximation of $\mu_t$ could be obtained in two steps, namely,
\begin{itemize}
\item drawing $N$ i.i.d. samples $\{ \bar \theta_t^{(i)} \}_{1\le i \le N}$ from the posterior measure at time $t-1$, $\mu_{t-1}$, 
\item and then computing normalised importance weights proportional to the marginal likelihoods $u_t(\bar \theta_t^{(i)})$. 
\end{itemize}
Unfortunately, neither sampling from $\mu_{t-1}$ nor the computation of the likelihood $u_t(\theta)$ can be carried out exactly, hence some approximations are in order.

\subsection{Jittering} \label{ssJittering}

Let us consider the problem of sampling first. Assume that a particle approximation $\mu_{t-1}^N = \frac{1}{N}\sum_{i=1}^N \delta_{\theta_{t-1}^{(i)}}$ of $\mu_{t-1}$ is available. In order to track the variations in $\mu_t$, it is convenient to have a procedure to generate a new set $\{ \bar \theta_{t-1}^{(i)} \}_{1 \le i \le N}$ which still yields an approximation of $\mu_{t-1}$ similar to $\mu_{t-1}^N$. A simple and practically appealing way to generate the new samples is to mutate the particles $\theta_{t-1}^{(1)}, ..., \theta_{t-1}^{(N)}$ 
independently using a {\em jittering} kernel $\kappa_N : \mB(D_\theta) \times D_\theta \rw [0,1]$, that we denote as 
\begin{equation}
\kappa_N(d\theta | \theta_{t-1}^{(i)}) = \kappa_N^{\theta_{t-1}^{(i)}}(d\theta), \quad i=1, 2, \ldots, N.
\label{eqKappa}
\end{equation}
The subscript $N$ in $\kappa_N$ indicates that the kernel may depend on the sample size $N$. This is a key feature in order to keep the distortion introduced by this mutation step sufficiently small, as will be made explicit in Section \ref{sConvergence} (see also Section \ref{ssJitter}).

\subsection{Conditional bootstrap filter and marginal likelihoods} \label{ssApproxLkd}

Let $\bar \theta_t^{(i)}$ be a Monte Carlo sample from $\kappa_N(d\theta | \theta_{t-1}^{(i)})$, i.e., a random mutation of $\theta_{t-1}^{(i)}$ as described above. The likelihood $u_t(\bar\theta_t^{(i)})$ can be approximated using Algorithm \ref{alBF} (the standard particle filter), conditional on $\Theta=\bar\theta_t^{(i)}$. For notational convenience, we introduce two random transformations of discrete sample sets on $\Real^{d_x}$, that will later be used to write down the conditional bootstrap filter.
\begin{Definicion}
Let $\{ x^{(j)} \}_{1 \le j \le M}$ be a set of $M$ points on the state space $\Real^{d_x}$. The set 
$$\{ \bar x^{(j)} \}_{1 \le j \le M} = \Upsilon_{n,\theta}\left( \{ x^{(j)} \}_{1 \le j \le M} \right)$$ 
is obtained by sampling each $\bar x^{(j)}$ from the corresponding transition kernel $\tau_{n,\theta}(dx|x^{(j)})$, for $j = 1, ..., M$.
\end{Definicion}
\begin{Definicion}
Let $\{ \bar x^{(j)} \}_{1 \le j \le M}$ be a set of $M$ points on the state space $\Real^{d_x}$. The set 
$$\{ x^{(j)} \}_{1 \le j \le M} = \Upsilon_{n,\theta}^{y_n}\left( \{ \bar x^{(j)} \}_{1 \le j \le M} \right)$$ 
is obtained by 
	\begin{itemize}
	\item computing normalised weights proportional to the likelihoods, 
	$$
	v_n^{(j)} = \frac{
		g_{n,\theta}^{y_n}(\bar x_n^{(j)})
	}{
		\sum_{k=1}^M g_{n,\theta}^{y_n}(\bar x_n^{(k)})
	}, \quad j=1, ..., M.
	$$
	\item and then resampling with replacement the set $\{ \bar x^{(j)} \}_{1 \le j \le M}$ according to the weights $\{ v_n^{(j)} \}_{1\le j\le M}$, i.e., assigning $x^{(j)} = \bar x^{(k)}$ with probability $v^{(k)}$, for $j=1, ..., M$ and $k \in \{ 1, ..., M \}$.
	\end{itemize}
\end{Definicion}

Let us now rewrite the bootstrap filter algorithm using this new notation.

\begin{Algoritmo} \label{alConditionalBF}
Bootstrap filter conditional on $\Theta = \theta_t^{(i)}$.
\begin{enumerate}
\item {\sf Initialisation.} Draw $M$ i.i.d. samples $x_0^{(i,j)}$, $j=1, ..., M$, from the prior distribution $\tau_0$.
\item {\sf Recursive step.} Let $\{ x_{n-1}^{(i,j)} \}_{1 \le j \le M}$ be the set of available samples at time $n-1$, with $n \le t$. The particle set is updated at time $n$ in two steps:
	\begin{enumerate}
	\item Compute $\{ \bar x_n^{(i,j)} \}_{1 \le j \le M} = \Upsilon_{n,\theta_t^{(i)}}\left( \{ x_{n-1}^{(i,j)} \}_{1 \le j \le M} \right)$.
	\item Compute $\{ x_n^{(i,j)} \}_{1 \le j \le M} = \Upsilon_{n,\theta_t^{(i)}}^{y_n}\left( \{ \bar x_n^{(i,j)} \}_{1 \le j \le M} \right)$.
	\end{enumerate}
\end{enumerate}
\end{Algoritmo}

For $n=t$, we obtain approximations of the posterior measures $\xi_{t,\bar \theta_t^{(i)}}(dx_t)$ and $\phi_{t,\bar\theta_t^{(i)}}(dx_t)$ of the form 
\begin{equation}
\xi_{t,\bar \theta_t^{(i)}}^M(dx_t) = \frac{1}{M} \sum_{j=1}^M \delta_{\bar x_t^{(i,j)}}(dx_t) \quad \mbox{and} \quad
\phi_{t,\bar\theta_t^{(i)}}^M(dx_t) = \frac{1}{M} \sum_{j=1}^M \delta_{x_t^{(i,j)}}(dx_t),
\label{eqXiPhi}
\end{equation} 
respectively, hence the likelihood $u_t(\bar\theta_t^{(i)})$ can be approximated as 
\begin{equation}
u_t^M(\bar\theta_t^{(i)}) = (g_{t,\bar\theta_t^{(i)}}^{y_t}, \xi_{t,\bar\theta_t^{(i)}}^M) 
= \frac{1}{M} \sum_{j=1}^M g_{t,\bar \theta_t^{(i)}}^{y_t}(\bar x_t^{(i,j)}).
\label{eqApproxLikelihood1}
\end{equation}


\subsection{Recursive algorithm}

If a new sample $\theta_t^{(i)} \in D_\theta$ is produced at time $t$, one can approximate the likelihood $u_t^M(\bar \theta_t^{(i)}) =  (g_{t,\bar\theta_t^{(i)}}^{y_t}, \xi_{t,\bar\theta_t^{(i)}}^M)$ by running a standard particle filter from time $0$ to time $t$, as shown in Section \ref{ssApproxLkd}.
However, the computational cost of this procedure obviously increases with time. We need to avoid this limitation in order to design a recursive algorithm.

Let us assume that the optimal filters $\phi_{t,\theta}(dx)$ are continuous w.r.t the parameter $\theta$, i.e., that if we have two candidate parameters $\theta$ and $\tilde \theta$ such that $\theta \approx \tilde \theta$, then $\phi_{t-1,\theta} \approx \phi_{t-1,\tilde \theta}$. If the latter approximation holds, then we can naturally expect that the predictive measure at time $t$ for the parameter $\tilde \theta$, namely $\xi_{t,\tilde \theta}$, can also be approximated using $\phi_{t-1,\theta}$ instead of $\phi_{t-1,\tilde \theta}$. To be specific, we can expect that
$$
\xi_{t,\tilde \theta} = \tau_{t,\tilde \theta}\phi_{t-1,\tilde\theta} \approx \tau_{t,\tilde\theta}\phi_{t-1,\theta}
$$
and, hence, the likelihood of the parameter $u_t(\tilde \theta) = (g_{t,\tilde\theta}^{y_t},\xi_{t,\tilde\theta})$, can be approximated from the filter at time $t-1$ computed for the {\em mismatched} parameter value $\theta$ (instead of the {\em actual} $\tilde\theta$), i.e.,
\begin{equation}
u_t(\tilde \theta) = (g_{t,\tilde\theta}^{y_t},\xi_{t,\tilde\theta}) \approx (g_{t,\tilde\theta}^{y_t},\tau_{t,\tilde\theta} \phi_{t-1,\theta}).
\label{eqSF0}
\end{equation}

If we accept the approximation in Eq. \eqref{eqSF0}, then it is possible to devise a truly recursive particle filter for the approximation of the posterior probability measures $\mu_t(d\theta)$. Assume that, at time $t-1$, we have been able to generate a set of particles in the parameter space $\{ \theta_{t-1}^{(i)} \}_{1 \le i \le N}$ and, for each $\theta_{t-1}^{(i)}$, we have the set of particles in the state space $\{ x_{t-1}^{(i,j)} \}_{1 \le j \le M}$. The latter set yields an approximation of the optimal filter conditional on $\theta_{t-1}^{(i)}$, i.e., we have 
\begin{equation}
\phi_{t-1,\theta_{t-1}^{(i)}} \approx \phi_{t-1,\theta_{t-1}^{(i)}}^M = \frac{1}{M} \sum_{j=1}^M \delta_{x_{t-1}^{(i,j)}}.
\nonumber
\end{equation} 
Now we generate a new parameter sample $\bar \theta_t^{(i)}$ by jittering the previous sample $\theta_{t-1}^{(i)}$ in a {\em controlled} manner (as suggested in Section \ref{ssJittering}). If the modulus of the difference, $\| \bar \theta_t^{(i)} - \theta_{t-1}^{(i)} \|$, is small enough, then we can expect that
\begin{equation}
\phi_{t-1,\bar \theta_t^{(i)}} \approx \phi_{t-1,\theta_{t-1}^{(i)}} \approx \phi_{t-1,\theta_{t-1}^{(i)}}^M = \frac{1}{M} \sum_{j=1}^M \delta_{x_{t-1}^{(i,j)}},
\label{eqSF1}
\end{equation}
i.e., we can use the particle approximation of the filter computed for $\theta_{t-1}^{(i)}$ as a particle approximation of the filter for the new sample $\bar \theta_t^{(i)}$. Once we have this approximation, it is straightforward to sample from the Markov kernels $\tau_{t,\bar \theta_t^{(i)}}(dx_t|x_{t-1}^{(i,j)})$ (this is the transformation $\Upsilon_{n,\bar \theta_t^{(i)}}$ applied to the set $\{ x_{t-1}^{(i,j)} \}_{1 \le j \le M}$ from which $\phi_{t-1,\theta_{t-1}^{(i)}}^M$ is constructed) in order to obtain the new predictive measure $\xi_{t,\bar \theta_t^{(i)}}^M$ and then approximate the likelihood of $\bar \theta_t^{(i)}$ as $u_t^M(\bar \theta_t^{(i)}) = (g_{t,\bar\theta_t^{(i)}}^{y_t},\xi_{t,\bar\theta_t^{(i)}}^M)$. In this process, we do not need to run a new particle filter from scratch, but simply to take a recursive step at time $t$. The price to pay is the introduction of an additional approximation error, that arises from \eqref{eqSF1} and needs to be quantified.

The complete recursive algorithm for the particle approximation of the sequence of measures $\mu_t$ is described below.


\begin{Algoritmo} \label{alRA}
Nested particle filtering for the approximation of $\mu_t$, $t=0, 1, 2, ...$
\begin{enumerate}
\item {\sf Initialisation.} Draw $N$ i.i.d. samples $\{ \theta_0^{(i)} \}_{1 \le i\le N}$ from the prior distribution $\pi_0(d\theta)$ and $N \times M$ i.i.d. samples $\{ x_0^{(i,j)} \}_{1 \le i \le N; 1 \le j \le M}$ from the prior distribution $\tau_0$. 

\item {\sf Recursive step.} For $t \ge 1$, assume the particle set $\left\{ \theta_{t-1}^{(i)}, \{ x_{t-1}^{(i,j)} \}_{1 \le j \le M} \right\}_{1 \le i \le N}$
is available and update it taking the following steps.

	\begin{itemize}
	\item[(a)] For each $i=1, ..., N$
		\begin{itemize}
		\item draw $\bar \theta_t^{(i)}$ from  $\kappa_N^{\theta_{t-1}^{(i)}}(d\theta)$,
		\item update $\{ \bar x_t^{(i,j)} \}_{1 \le j \le M} = \Upsilon_{t,\bar \theta_t^{(i)}}\left(
			\{ x_{t-1}^{(i,j)} \}_{1 \le j \le M}
		\right)$ and construct $\xi_{t,\bar \theta_t^{(i)}}^M = \frac{1}{M}\sum_{j=1}^M \delta_{\bar x_t^{(i,j)}}$,
		\item compute the approximate likelihood $u_t^M(\bar \theta_t^{(i)}) = ( g_{t,\bar \theta_t^{(i)}}^{y_t}, \xi_{t,\bar\theta_t^{(i)}}^M )$, and
		\item update the particle set $\{ \tilde x_t^{(i,j)} \}_{1 \le j \le M} = \Upsilon_{t,\bar \theta_t^{(i)}}^{y_t}\left(
			\{ \bar x_{t}^{(i,j)} \}_{1 \le j \le M}
		\right)$.
		\end{itemize}
	\item[(b)] Compute normalised weights $w_t^{(i)} \propto u_t^M(\bar \theta_t^{(i)})$, $i=1, ..., N$.
	\item[(c)] Resample: for each $i=1, ..., N$, set $\left\{ \theta_t^{(i)},  x_t^{(i,j)} \right\}_{ 1 \le j \le M} = \left\{ \bar \theta_t^{(l)}, \tilde x_t^{(l,j)} \right\}_{1 \le j \le M}$ with probability $w_t^{(l)}$, where $l \in \{ 1, ..., N \}$.
	\end{itemize}
\end{enumerate}
\end{Algoritmo}

Step 2(a) in Algorithm \ref{alRA} involves jittering the samples in the parameter space and then taking a single recursive step of a bank of $N$ standard particle filters. In particular, for each $\bar \theta_t^{(i)}$, $1 \le i \le N$, we have to propagate the particles $\{ x_{t-1}^{(i,j)} \}_{1 \le j \le M}$ so as to obtain a new set $\{ \tilde x_t^{(i,j)} \}_{1 \le j \le M}$. 

\begin{Nota}
The cost of the recursive step in Algorithm \ref{alRA} is independent of $t$. We only have to carry out regular `prediction' and `update' operations in a bank of standard particle filters. Hence, Algorithm \ref{alRA} is sequential, purely recursive and can be implemented online.
\end{Nota} 

\begin{Nota}
Algorithm \ref{alRA} yields several approximations. While $\mu_t^{N,M}=\frac{1}{N}\sum_{i=1}^N \delta_{\theta_t^{(i)}}$ is an estimate of $\mu_t$, the joint posterior measure $\pi_t$ is approximated as $\pi_t^{N,M} = \frac{1}{NM} \sum_{i=1}^N\sum_{j=1}^M \delta_{\theta_t^{(i)},x_t^{(i,j)}}$. Conditional predictive and filter measures on the state space are also computed by the inner filters, namely
$
\xi_{t,\bar \theta_t^{(i)}}^M = \frac{1}{M}\sum_{j=1}^M \delta_{\bar x_t^{(i,j)}}
$
and 
$
\phi_{t,\theta_t^{(i)}}^M = \frac{1}{M}\sum_{j=1}^M \delta_{x_t^{(i,j)}}.
$ 
\end{Nota}

\section{Summary of results} \label{sResults}


\subsection{Convergence of the approximation errors in $L_p$}

We pursue a characterisation of the $L_p$ norms of the approximation errors for $\mu_t^{N,M}$, $\phi_{t,\theta_t^{(i)}}^M$ ($i=1, ..., N$) and $\pi_t^{N,M}$ which can be stated in a form similar to Lemma \ref{lmUnifConvPF}. Towards this aim, we prove in Section \ref{sConvergence} that, under regularity assumptions on the state-space model and the jittering kernel $\kappa_N^\theta$, the $L_p$ norms of the errors asymptotically decrease toward 0, and provide explicit convergence rates. To be specific, our analysis relies on the following basic assumptions (to be stated in a precise manner in Section \ref{sConvergence}):
\begin{itemize}
\item The optimal filters $\phi_{t,\theta}$ are continuous w.r.t. the parameter $\theta$.
\item The jittering steps are ``small enough'' and, in particular, the variance of the jittering kernel is a decreasing function of the number of particles $N$.
\item The parameter $\theta$ is restricted to take values on a compact set $D_\theta$, and the conditional pdf of the observations, $g_{t,\theta}^{y_t}(x_t)$ is positive and uniformly bounded over $D_\theta$.
\end{itemize}
The continuity of the optimal filters and the constraint on the variance of the jittering kernel are at the core of Algorithm \ref{alRA}. If these conditions are not satisfied, it cannot be expected to converge, as the errors due to the jittering steps may grow without bound. Under the assumptions above, we have proved the results below, that hold true for an arbitrary-but-fixed sequence of observations $y_{1:T}$, with $T<\infty$, and arbitrary test functions $h\in B(D_\theta)$ and $f\in B(D_\theta \times \Real^{d_x})$.

\begin{Result} \label{reMu}
(Theorem \ref{thConvergenceLp}, Section \ref{sConvergence}). 
There exist constants $c_t, \bar c_t < \infty$, independent of $N$ and $M$, such that
$$
\| (h,\mu_t^{N,M}) - (h,\mu_t) \|_p \le \frac{c_t\|h\|_\infty}{\sqrt{N}} + \frac{\bar c_t\|h\|_\infty}{\sqrt{M}}
$$ 
for any $p\ge 1$ and every $t = 0, \ldots, T$.
\end{Result}

\begin{Result} \label{rePi}
(Theorem \ref{thConvergenceL1-Joint}, Section \ref{sConvergence}).
There exist constants $c_t,\bar c_t<\infty$, independent of $N$ and $M$, such that
$$
\| (f,\pi_t^{N,M}) - (f,\pi_t) \|_p \le \frac{c_t\|h\|_\infty}{\sqrt{N}} + \frac{\bar c_t\|h\|_\infty}{\sqrt{M}}
$$ 
for any $p\ge 1$ and every $t = 0, \ldots, T$.
\end{Result}

Additionally, Algorithm \ref{alRA} yields explicit approximations of the conditional filter measures (for $\Theta = \theta_t^{(i)}$, $i=1, ..., N$). In particular, we will show that the statement below also holds under mild assumptions.

\begin{Result}
(Remark \ref{rmConvergenceInnerFilters}, Section \ref{sConvergence}).
For any $l \in B(\Real^{d_x})$ there exist constants $k_t,\bar k_t<\infty$, independent of $M$ and $N$, such that
$$
\sup_{1 \le i \le N} \| (l,\phi_{t,\theta_t^{(i)}}^M) - (l,\phi_{t,\theta_t^{(i)}}) \|_p \le \frac{k_t\|l\|_\infty}{\sqrt{N}} + \frac{\bar k_t\|l\|_\infty}{\sqrt{M}}
$$ 
for any $p\ge 1$ and every $t = 0, \ldots, T$. 
\end{Result}

\begin{Nota} \label{rmMuPi}
In most practical applications we can expect constraints on the computational effort that can be invested at each time step. Typically, this occurs because a full sequential step of the algorithm must be completed before a new observation is received. This is likely to impose a limitation on the overall number of samples that can be generated, namely the product $K=MN$. For a given value of $K$ (say with integer $\sqrt{K}$), Results \ref{reMu} and \ref{rePi} above indicate that the choice of $M$ and $N$ that minimises the error rate is $M=N=\sqrt{K}$. In this case, we obtain approximate measures
$$
\hat \mu_t^K \dfn \frac{1}{\sqrt{K}} \sum_{i=1}^{\sqrt{K}} \delta_{\theta_t^{(i)}} 
\quad \mbox{and} \quad 
\hat \pi_t^K \dfn \frac{1}{K} \sum_{i=1}^{\sqrt{K}} \sum_{j=1}^{\sqrt{K}} \delta_{\theta_t^{(i)},x_t^{(i,j)}}
$$
such that
$$
\| (h, \hat \mu_t^K) - (h,\mu_t) \|_p \le \frac{\hat c_t\|h\|_\infty}{K^{\frac{1}{4}}} 
\quad \mbox{and} \quad
\| (f, \hat \pi_t^K) - (f,\pi_t) \|_p \le \frac{\hat c_t\|f\|_\infty}{K^{\frac{1}{4}}}, 
$$
for any test functions $h \in B(D_\theta)$ and $f \in B(D_\theta \times \Real^{d_x})$, and some finite constants $\hat c_t$ and $\hat c_t$.
\end{Nota}

\subsection{Jittering} \label{ssJitter}

The main choice to be made when implementing the algorithm is the type of jittering kernel, as in Eq. \eqref{eqKappa}, to be used. This can actually be very simple. Assume for instance a standard Gaussian kernel $\hat \kappa^{\theta'}$, with mean $\theta'$ and covariance matrix $C={\mathcal I}_{d_\theta}$, where ${\mathcal I}_{d_\theta}$ is the $d_\theta \times d_\theta$ identity matrix, and let $\kappa^{\theta'}$ the corresponding kernel truncated within the parameter support set $D_\theta$. Any kernel of the form
\begin{equation}
\kappa_N^{\theta'} = (1-\epsilon_N) \delta_{\theta'} + \epsilon_N \kappa^{\theta'},
\label{eqKernelAux}
\end{equation}
with $\epsilon_N \le \frac{1}{N^{\frac{p}{2}}}$ is sufficient to make Results \ref{reMu} and \ref{rePi} hold with a prescribed value of $p$. Note that the choice of $\kappa_N$ in \eqref{eqKernelAux} amounts to perturbing each particle with probability $\epsilon_N$ (or leave it unchanged with probability $1-\epsilon_N$). The perturbations applied can be large, but not many particles are actually perturbed. 

Alternatively, we can choose a standard Gaussian kernel $\hat \kappa_N^{\theta'}$, with mean $\theta'$ and covariance matrix $C_N \propto \frac{1}{N^\frac{p+2}{p}} {\mathcal I}_{d_\theta}$. The jittering kernel $\kappa_N^{\theta'}$ is then obtained by truncating $\hat \kappa_N^{\theta'}$ within the parameter support set $D_\theta$. In this case we perturb every particle, but each single perturbation is small. This choice of $\kappa_N$ is also sufficient for Results \ref{reMu} and \ref{rePi} to hold. See Section \ref{ssJitterAnalysis} and Appendix \ref{apPequeLema} for a detailed description. 




In practice, the magnitude of the jittering introduced by the kernel $\kappa_N$ is relevant for the performance of the algorithm, because it determines how fast the support of the approximating measure $\mu_t^{N,M}$ can be adapted over time to track changes\footnote{The jittering step enables the adaptation of the support set $\{ \theta_t^{(i)} \}_{1 \le N}$. The {\em shape} of the posterior distribution is tracked by computing the importance weights.}. If the jittering variance is too small, it may turn out hard to track large changes in the posterior measure $\mu_t$. Such large changes can be expected for small $t$ (when the amount of accumulated data is still limited), in the presence of outliers, due to change-points not accounted for by the model, etc. Some specific techniques can be adapted from \citep{Maiz12} to deal with outliers, and we show a simple numerical example at the end of Section \ref{sExamples} to illustrate the effect of change-points. On the other hand, if the jittering variance is made too large, the adaptivity of the algorithm can be improved but its converge rate can be compromised (see Remark \ref{rmSlowConv} in Section \ref{ssWeights}).

\subsection{Comparison with the SMC$^2$ method} \label{ssComparison}

The natural benchmark for the algorithm introduced in this paper is the SMC$^2$ method of \citep{Chopin12}. This technique is similar in structure to Algorithm \ref{alRA} and, in particular, it generates and maintains over time $N$ particles in the parameter space and, for each one of them, $M$ particles in the state space. However, it displays two key differences w.r.t. Algorithm \ref{alRA}:
\begin{itemize}
\item The particles in the parameter space are jittered using a particle MCMC kernel, with the aim of leaving the approximate posterior distribution of the parameters invariant.
\item The weights for the particles in the parameter space at time $t$ are computed using the complete sequence of observations $y_{1:t}$. 
\end{itemize}
The SMC$^2$ algorithm is consistent  \cite[Proposition 1]{Chopin12}, as it targets a sequence of probability measures (of increasing dimension) that have the parameter posterior measures, $\{ \mu_t \}_{t\ge 0}$, as marginals. Although this is not expicitly proved in \citep{Chopin12}, under adequate assumptions it can be shown that the SMC$^2$ method produces approximate measures $\mu_{t,SMC}^{N,M}$ such that the $L_p$ norms of the approximation errors can be bounded as 
\begin{equation}
\| (h,\mu_{t,SMC}^{N,M} - (h,\mu_t)\|_p \le \frac{C_t}{\sqrt{N}}
\label{eqSF10}
\end{equation}
for some constant $C_t$, independent of $N$ and $M$. This implies that the approximation errors vanish asymptotically as $N\rw\infty$, even if $M<\infty$ is kept fixed. Also, if $K=NM$ is the total number of particles in the state space generated by the SMC$^2$ algorithm, and $M$ is assumed to be constant, the the inequality \eqref{eqSF10} implies that the approximation errors converge as $K^{-\frac{1}{2}}$.

The obvious drawback of the SMC$^2$ method is that it is not recursive: {\em both} the use of a particle MCMC kernel\footnote{Particularly note that if we replace the jittering kernel in the proposed Algorithm \ref{alRA} by a particle MCMC kernel, the resulting procedure is not recursive anymore.} {\em and} the computation of the particle weights at time $t$ involve the processing of the whole sequence of observations $y_{1:t}$. In particular, a straightforward implementation of the SMC$^2$ algorithm with periodic resampling steps and a sequence of $T$ observations, $y_{1:T}$, yields complexity $O(NMT^2)$. In comparison, Algorithm \ref{alRA} is purely recursive, hence for a sequence of observations $y_{1:T}$ the computational cost is $O(NMT)$, i.e., linear in $T$ versus the quadratic complexity of the original SMC$^2$ approach.


The linear complexity $O(NMT)$ of Algorithm \ref{alRA}, however, comes at the expense of some limitations compared to the SMC$^2$ technique. The most important one is that the approximation errors converge with $\frac{1}{\sqrt{N}} + \frac{1}{\sqrt{M}}$ (see Result \ref{reMu}), hence we need to let $N \rw \infty$ and $M \rw \infty$ for the errors to vanish, while in the SMC$^2$ method it is enough to have $N\rw\infty$ (and keep $M$ fixed). If $K=NM$ is the total number of particles in the state space, the optimal allocation for Algorithm \ref{alRA} is $N=M=\sqrt{K}$ and the convergence rate is $K^{-\frac{1}{4}}$ (see Remark \ref{rmMuPi}) while the SMC$^2$ attains a rate $K^{-\frac{1}{2}}$.

We finally remark that the conditional optimal filters $\phi_{t,\theta}$ need to be continuous w.r.t. $\theta \in D_\theta$ in order to ensure the convergence of Algorithm \ref{alRA}, while this is not necessary for the SMC$^2$, the particle MCMC \citep{Andrieu10} or the nonlinear population Monte Carlo \citep{Koblents15} methods. This limitation of the proposed scheme is a direct consequence of not using the full sequence of observations to compute the weights.

%

\section{Convergence analysis} \label{sConvergence}

We split the analysis of the recursive Algorithm \ref{alRA} in three steps: jittering, weight computation and resampling. At the beginning of time step $t$, the approximation $\mu_{t-1}^{N,M}$ of $\mu_{t-1}$ is available. After the jittering step we have a new approximation, 
$$
\bar \mu_{t-1}^{N,M} = \frac{1}{N}\sum_{i=1}^N \delta_{\bar \theta_t^{(i)}},
$$ 
and we need to prove that it converges to $\mu_{t-1}$. After the computation of the weights, the measure 
$$
\tilde \mu_t^{N,M} = \sum_{i=1}^N w_t^{(i)} \delta_{\bar \theta_t^{(i)}}
$$
is obtained (note that the weights $w_t^{(i)} \propto \left( g_{t,\bar \theta_t^{(i)}}^{y_t},\xi_{t,\bar \theta_t^{(i)}}^M \right)$ depend on $M$, although we skip this dependence for notational simplicity) and its convergence toward $\mu_t$ must be established. Finally, after the resampling step, we need to prove that 
$$
\mu_t^{N,M} = \frac{1}{N} \sum_{i=1}^N \delta_{\theta_t^{(i)}}
$$
converges to $\mu_t$ in an appropriate manner. We prove the convergence of $\bar \mu_{t-1}^N$, $\tilde \mu_t^N$ and $\mu_t^N$ in three corresponding lemmas and then combine them to prove the asymptotic convergence of Algorithm \ref{alRA}. 
Splitting the proof has the advantage that we can ``reuse'' these partial lemmas easily in order to prove different statements. For example, it is straightforward to show that $\pi_t^{N,M} \rw \pi_t$, when $N,M \rw \infty$, as well (see Section \ref{ssJoint}).


\subsection{Jittering step} \label{ssJitterAnalysis}

In the jittering step, a rejuvenated cloud of particles is generated by propagating the existing samples across the kernels $\kappa_N^{\theta_{t-1}^{(i)}}$, $i=1, ..., N$. For the analysis, we abide by the following assumption.

\begin{Assumption} \label{asKernel}
The family of kernels $\kappa_N^{\theta'}$, $\theta' \in D_\theta$, used in the jittering step satisfy the inequality 
\begin{equation}
\sup_{\theta' \in D_\theta} \int | h(\theta) - h(\theta') | \kappa_N^{\theta'}(d\theta) \le \frac{
	c_\kappa\| h \|_\infty
}{
	\sqrt{N}
}
\label{eqAsKernel1}
\end{equation}
for any $h \in B(D_\theta)$ and some constant $c_\kappa < \infty$.
\end{Assumption}

\begin{Nota} \label{rmKernel-1}
One simple class of kernels that complies with A.\ref{asKernel} has the form  
\begin{equation}
\kappa_N^{\theta'}(d\theta) = (1-\epsilon_N)\delta_{\theta'}(d\theta) + \epsilon_N \bar \kappa_N^{\theta'}(d\theta),
\label{eqExampleKappa1}
\end{equation}
where $0 \le \epsilon_N \le \frac{1}{\sqrt{N}}$ and $\bar \kappa_N^{\theta'} \in \mP(D_\theta)$ for every $\theta' \in D_\theta$. Note that substituting \eqref{eqExampleKappa1} into \eqref{eqAsKernel1} yields
\begin{equation}
\sup_{\theta' \in D_\theta} \int | h(\theta) - h(\theta') | \kappa_N^{\theta'}(d\theta) \le 2 \epsilon_N \| h \|_\infty \le \frac{
	2\| h \|_\infty
}{
	\sqrt{N}
},
\nonumber
\end{equation}
hence A.\ref{asKernel} is satisfied with $c_\kappa = 2$.

When using a kernel of the form in \eqref{eqExampleKappa1} only a small fraction of particles are actually changed in the jittering step. However, when a particle is actually jittered, the move can be large. Note that the variance of $\bar \kappa_N^{\theta'}(d\theta)$ can be independent of $N$ and possibly large, since the variance of $\kappa_N^{\theta'}(d\theta)$ is controlled by the choice of $\epsilon_N \le \frac{1}{\sqrt{N}}$ alone.
\end{Nota}

\begin{Nota} \label{rmKernel-2}
Assume that $h \in B(D_\theta)$ is Lipschitz, i.e., there is a constant $c_L < \infty$ such that 
\begin{equation}
| h(\theta) - h(\theta') | \le c_L\| h \|_\infty \| \theta - \theta' \|
\nonumber
\end{equation}
for any $\theta, \theta' \in D_\theta$. If there exists a constant $\breve c < \infty$ independent of $N$ such that the inequality
\begin{equation}
\sigma_{\kappa,N}^2 = \sup_{\theta'\in D_\theta} \int \| \theta - \theta' \|^2 \kappa_N^{\theta'}(d\theta) \le \frac{
	\breve c
}{
	\epsilon_N^3 N^\frac{3}{2}
}
\label{eqIneqSigmaKappa}
\end{equation}
is satisfied, then Eq. \eqref{eqAsKernel1} in A.\ref{asKernel} holds with $c_\kappa = c_L\left(
	1 + \breve c \sup_{\theta_1,\theta_2 \in D_\theta} \| \theta_1 - \theta_2 \|
\right) < \infty$. A generalization of this statement is proved in Appendix \ref{apPequeLema}. Note that with this class of kernels every particle is jittered at each time step, but the moves are very small.

\end{Nota}

\begin{Lema} \label{lmSampling}
Let $Y_{1:T}=y_{1:T}$ be arbitrary but fixed and choose any $0<t\le T$. If $h \in B(D_\theta)$, A.\ref{asKernel} holds and
\begin{equation}
\| (h,\mu_{t-1}^{N,M}) - (h,\mu_{t-1}) \|_p \le \frac{c_{t-1}\|h\|_\infty}{\sqrt{N}} + \frac{\bar c_{t-1}\|h\|_\infty}{\sqrt{M}}
\label{eqHypoLemmaKernel}
\end{equation} 
for some $p \ge 1$ and some constants $c_{t-1}, \bar c_{t-1} < \infty$ independent of $N$ and $M$, then 
\begin{equation}
\| (h,\bar \mu_{t-1}^{N,M}) - (h,\mu_{t-1}) \|_p \le \frac{c_{1,t}\|h\|_\infty}{\sqrt{N}} + \frac{\bar c_{1,t}\|h\|_\infty}{\sqrt{M}},
\label{eqLemmaKernel}
\end{equation}
where the constants $c_{1,t}, \bar c_{1,t} < \infty$ are also independent of $N$ and $M$.
\end{Lema}

\noindent {\bf Proof:}  Recall that we draw the particles $\bar \theta_t^{(i)}$, $i=1, \ldots, N$, independently from the  kernels $\kappa_N^{\theta_{t-1}^{(i)}}$, $i=1, \ldots, N$, respectively. In order to prove that \eqref{eqLemmaKernel} holds, we start from the iterated triangle inequality 
\begin{eqnarray}
\| (h,\bar \mu_{t-1}^{N,M}) - (h,\mu_{t-1}) \|_p &\le& 
\| (h,\bar \mu_{t-1}^{N,M}) - (h, \kappa_N\mu_{t-1}^{N,M}) \|_p \nonumber\\ 
&& + \| (h,\kappa_N\mu_{t-1}^{N,M}) - (h, \mu_{t-1}^{N,M}) \|_p \nonumber\\
&& + \| (h, \mu_{t-1}^{N,M}) - (h,\mu_{t-1}) \|_p,
\label{eqRetriangle}
\end{eqnarray}
where
$$
(h,\kappa_N\mu_{t-1}^{N,M}) = \frac{1}{N} \sum_{i=1}^N \int h(\theta) \kappa_N^{\theta_{t-1}^{(i)}}(d\theta),
$$
and then analyse each of the terms on the right hand side of \eqref{eqRetriangle} separately. Note that the last term, in particular, is straightforward: its bound follows directly from the assumption in Eq. \eqref{eqHypoLemmaKernel}.

Let $\mG_{t-1}$ be the $\sigma$-algebra generated by the random particles 
$\{ \bar \theta_{1:t-1}^{(i)}, \theta_{0:t-1}^{(i)} \}_{1 \le i \le N}$. 
Then 
\begin{equation}
E\left[
	(h,\bar \mu_{t-1}^{N,M}) | \mG_{t-1}
\right] = 
\frac{1}{N} \sum_{i=1}^N \int h(\theta)\kappa_N^{\theta_{t-1}^{(i)}}(d\theta) = (h,\kappa_N\mu_{t-1}^{N,M})
\nonumber
\end{equation}
and the difference $(h,\bar \mu_{t-1}^{N,M}) - (h,\kappa_N\mu_{t-1}^{N,M})$ can be written as
\begin{equation}
(h,\bar \mu_{t-1}^{N,M}) - (h,\kappa_N\mu_{t-1}^{N,M}) = \frac{1}{N} \sum_{i=1}^N \bar Z_{t-1}^{(i)},
\nonumber
\end{equation}
where the random variables $\bar Z_{t-1}^{(i)} = h(\bar \theta_t^{(i)}) - E[ h(\bar \theta_t^{(i)}) | \mG_{t-1} ]$, $i=1, ..., N$, are conditionally independent (given $\mG_{t-1}$), have zero mean and can be bounded as $| \bar Z_{t-1}^{(i)} | \le 2\| h \|_\infty$. It is an exercise in combinatorics to show that the number of non-zero terms in 
$$
E\left[ \left( \sum_{i=1}^N \bar Z_{t-1}^{(i)} \right)^p \left| \mG_{t-1} \right. \right] = \sum_{i_1}\cdots\sum_{i_p} E\left[
	\bar Z_{t-1}^{(i_1)} \ldots \bar Z_{t-1}^{(i_p)}
	\left| \mG_{t-1} \right.
\right]
$$ 
is a polynomial of order no greater than $N^{\frac{p}{2}}$ with coefficients independent of $N$. As a consequence, there exists a constant $\tilde c_1$, independent of $N$, $M$ and $h$ (actually independent of the distribution of the $\bar Z_{t-1}^{(i)}$'s) such that
\begin{equation}
E\left[
	\left|
		(h,\bar \mu_{t-1}^{N,M}) - (h,\kappa_N\mu_{t-1}^{N,M})
	\right|^p | \mG_{t-1}
\right] = E\left[
	\left|
		\frac{1}{N} \sum_{i=1}^N \bar Z_{t-1}^{(i)}
	\right|^p | \mG_{t-1}
\right] \le \frac{
	\tilde c_{1}^p \| h \|_{\infty}^p
}{
	N^\frac{p}{2}
}. \label{eqBoundingFirstTerm-1}
\end{equation}
From \eqref{eqBoundingFirstTerm-1} we readily obtain that
\begin{equation}
\| (h,\bar \mu_{t-1}^{N,M}) - (h,\kappa_N\mu_{t-1}^{N,M}) \|_p \le \frac{
	\tilde c_{1}\| h \|_\infty
}{
	\sqrt{N}
}.
\label{eqBoundingFirstTerm-2}
\end{equation}

For the remaining term in \eqref{eqRetriangle}, namely, $\| (h,\kappa_N\mu_{t-1}^{N,M}) - (h,\mu_{t-1}^{N,M})\|_p$, we simply note that
\begin{eqnarray}
\left| 
	(h,\kappa_N\mu_{t-1}^{N,M}) - (h,\mu_{t-1}^{N,M})
\right| &=& \left|
	\frac{1}{N} \sum_{i=1}^N \int \left(
		h(\theta) - h(\theta_{t-1}^{(i)})
	\right) \kappa_N^{\theta_{t-1}^{(i)}}(d\theta)
\right| \nonumber \\
&\le& \frac{1}{N} \sum_{i=1}^N \int \left|
	h(\theta) - h(\theta_{t-1}^{(i)})
\right| \kappa_N^{\theta_{t-1}^{(i)}}(d\theta) 
\le \frac{
	c_\kappa \|h\|_\infty
}{
	\sqrt{N}
}, \nonumber\\
\label{eqBoundingSecond-1}
\end{eqnarray}
where the last inequality follows from assumption A.\ref{asKernel}, with the constant $c_\kappa<\infty$ independent of $N$ and $M$. 


Substituting the inequalities \eqref{eqHypoLemmaKernel}, \eqref{eqBoundingFirstTerm-2} and \eqref{eqBoundingSecond-1} into Eq. \eqref{eqRetriangle} yields the desired conclusion, viz., Eq. \eqref{eqLemmaKernel}, with constants $c_{1,t} = c_{t-1} + c_\kappa + \tilde c_1$ and $\bar c_{1,t} = \bar c_{t-1}$ independent of $N$ and $M$.
$\QED$

\subsection{Computation of the weights} \label{ssWeights}

Since the integral $u_t(\theta) = (g_{t,\theta}^{y_t},\xi_{t,\theta})$ is intractable, the importance weights are computed as
$$
w_t^{(i)} \propto ( g_{t,\bar \theta_t^{(i)}}^{y_t}, \xi_{t,\bar \theta_t^{(i)}}^M ) = u_t^M(\bar \theta_t^{(i)}), \quad i=1, ..., N.
$$
We also recall that the particles in the set $\{ x_{t-1}^{(i,j)} \}_{1 \le j \le M}$, which yield the approximate filter $\phi_{t-1,\theta_{t-1}^{(i)} }^M = \frac{1}{M} \sum_{j=1}^M \delta_{x_{t-1}^{(i,j)}}$, are propagated through the transition kernels as
$$
\bar x_t^{(i,j)} \sim \tau_{t,\bar \theta_t^{(i)}}(dx_t|x_{t-1}^{(i)}), 
\quad 
j=1, \ldots, M, \quad \mbox{to obtain} 
\quad 
\xi_{t,\bar \theta_t^{(i)} }^M = \frac{1}{M} \sum_{j=1}^M \delta_{\bar x_t^{(i,j)}}.
$$
This means that we are using $\phi_{t-1,\theta_{t-1}^{(i)}}^M$ as an estimate of $\phi_{t-1,\bar \theta_t^{(i)}}$ in order to compute the predictive measure $\xi_{t,\bar \theta_t^{(i)}}^M$ and, as a consequence, it is necessary to prove that the error introduced at this step can be bounded in the same way as the approximation errors in Lemma \ref{lmSampling}. To attain that result, we need to strengthen slightly our assumptions on the structure of the kernel $\kappa_N$.

\begin{Assumption} \label{asKernel-2}
The family of kernels $\kappa_N^{\theta'}$, $\theta' \in D_\theta$, used in the jittering step satisfies the inequality 
\begin{equation}
\sup_{\theta' \in D_\theta} \int \| \theta - \theta' \|^p \kappa_N^{\theta'}(d\theta) \le \frac{
	c_\kappa^p
}{
	N^\frac{p}{2}
}
\label{eqAsKernel2}
\end{equation}
for some prescribed $p \ge 1$ and some constant $c_\kappa < \infty$.
\end{Assumption}

\begin{Nota} \label{rmCkappa}
It is simple to prove that kernels of the class 
\begin{equation}
\kappa_N^{\theta'} = (1-\epsilon_N)\delta_{\theta'} + \epsilon_N \bar \kappa_N^{\theta'},
\label{eqDefKernel}
\end{equation}
with $0<\epsilon_N\le \frac{1}{N^\frac{p}{2}}$ and $\bar \kappa_N^{\theta'} \in \mP(D_\theta)$, satisfy assumption A.\ref{asKernel-2} for every $p \ge 1$. Simply note that
\begin{equation}
\sup_{\theta' \in D_\theta} \int \| \theta - \theta' \|^p \kappa_N^{\theta'}(d\theta) \le \epsilon_N \hat C^p \le \frac{\hat C^p}{N^\frac{p}{2}},
\nonumber
\end{equation}
where $\hat C^p = \sup_{\theta_1,\theta_2\in D_\theta} \| \theta_1-\theta_2 \|^p < \infty$, since $D_\theta$ is compact. The inequality \eqref{eqAsKernel2} also holds for any kernel $\kappa_N^{\theta'}$ that satisfies the inequality
\begin{equation}
\sigma_{\kappa,N}^2 = \sup_{\theta,\theta' \in D_\theta} \int \| \theta - \theta' \|^2 \kappa_N^{\theta'}(d\theta) \le \frac{
	\breve c
}{
	N^\frac{p+2}{2}
}
\label{eqSigma2Kernel2}
\end{equation}
for some constant $\breve c < \infty$ (see Appendix \ref{apPequeLema} for a generalisation of this result). 

In the first case, $\epsilon_N \le \frac{1}{\sqrt{N}}$, we control the number of particles that are jittered. However, those which are actually jittered may experience large perturbations. In the second case, we allow for the jittering of all particles but, in exchange, the second order moment of the perturbation is controlled. Kernels of the class in \eqref{eqDefKernel} with $\epsilon_N \le \frac{1}{\sqrt{N}}$ trivially satisfy A.\ref{asKernel}. The inequality \eqref{eqAsKernel1} in A.\ref{asKernel} also holds for any kernel $\kappa_N^{\theta'}$ that satisfies \eqref{eqSigma2Kernel2} for the prescribed value of $p$.   
\end{Nota}

\begin{Nota}
\label{rmSlowConv}
It is possible to replace the factor $N^{-\frac{1}{2}}$ in assumptions A.\ref{asKernel} and A.\ref{asKernel-2} by some strictly decreasing function of $N$, say $r(N)$, and still prove the convergence of the nested particle filtering scheme (Algorithm \ref{alRA}). However, the error rates would depend directly on the choice of $r(N)$, so that if $r(N) > N^{-\frac{1}{2}}$, then convergence would be attained at a slower pace (relative to $N$). If $r(N)$ were chosen to be constant, convergence would not be guaranteed.
\end{Nota}

Using $\phi_{t-1,\theta_{t-1}^{(i)}}^M$ as an estimate of $\phi_{t-1,\bar \theta_t^{(i)}}$ can only work consistently if the filter measure $\phi_{t-1,\theta}$ is continuous in the parameter $\theta$. Here we assume that $\phi_{t-1,\theta}$ is Lipschitz, as stated below.

\begin{Assumption} \label{asPhiLipschitz}
The measures $\phi_{t,\theta}$, $t \ge 1$, are Lipschitz in the parameter $\theta \in D_\theta$. Specifically, for every function $f \in B(\Real^{d_x})$ there exists a constant $b_t < \infty$ such that
$$
| (f,\phi_{t,\theta'}) - (f,\phi_{t,\theta''}) | \le b_{t} \|f\|_\infty \| \theta' - \theta'' \| \quad \mbox{for any } \theta',\theta'' \in D_\theta.
$$  
\end{Assumption}

Assumptions A.\ref{asKernel-2} and A.\ref{asPhiLipschitz} enable us to quantify the error $\| (f,\phi_{t-1,\bar \theta_t^{(i)}}) - (f,\phi_{t-1,\theta_{t-1}^{(i)}}^M) \|_p$, as made explicit by the following lemma.

\begin{Lema} \label{lmRecursive}
Assume that:
\begin{itemize}
\item[(a)] A.\ref{asPhiLipschitz} holds (i.e., $\phi_{t-1,\theta}$ is Lipschitz in $\theta$);
\item[(b)] for any $\theta' \in D_\theta$ and $f \in B(\Real^{d_x})$, $\phi_{t-1,\theta'}^M$ is a random measure that satisfies the inequality
\begin{equation}
\| (f, \phi_{t-1,\theta'}^M) - (f,\phi_{t-1,\theta'}) \|_p \le \frac{
	c_{t-1}\|f\|_\infty
}{
	\sqrt{N}
} + \frac{
	\bar c_{t-1}\|f\|_\infty
}{
	\sqrt{M}
},
\nonumber
\end{equation}
for some constants $c_{t-1}, \bar c_{t-1} < \infty$ independent of $N$, $M$ and $\theta'$; and 
\item[(c)] the random parameter $\theta''$ is distributed according to a probability measure $\kappa_N^{\theta'}(d\theta)$ that complies with A.\ref{asKernel-2} for some prescribed $p \ge 1$.
\end{itemize}
Then, for every $f \in B(\Real^{d_x})$ and every $\theta' \in D_\theta$, there exist constants $\tilde c_{t-1}, \bar {\tilde c}_{t-1}< \infty$, independent of $N$, $M$ and $\theta'$, such that
$$
\| (f,\phi_{t-1,\theta'}^M) - (f,\phi_{t-1,\theta''}) \|_p \le \frac{
	\tilde c_{t-1}\|f\|_\infty
}{
	\sqrt{N}
} + \frac{
	\bar {\tilde c}_{t-1}\|f\|_\infty
}{
	\sqrt{M}
}.
$$
\end{Lema}

\noindent {\bf Proof:} Consider the triangle inequality
\begin{equation}
\| (f,\phi_{t-1,\theta'}^M) - (f,\phi_{t-1,\theta''}) \|_p \le \| (f,\phi_{t-1,\theta'}^M) - (f,\phi_{t-1,\theta'}) \|_p + \| (f,\phi_{t-1,\theta'}) - (f,\phi_{t-1,\theta''}) \|_p.
\label{eqTriangle-Distort}
\end{equation}
We aim at bounding the two terms on the right hand side of \eqref{eqTriangle-Distort}.

For the first term, we simply apply assumption (b) in the statement of Lemma \ref{lmRecursive}, which yields
\begin{equation}
\| (f,\phi_{t-1,\theta'}^M - (f,\phi_{t-1,\theta'}) \|_p \le \frac{
	c_{t-1}\|f\|_\infty
}{
	\sqrt{N}
} + \frac{
	\bar c_{t-1}\|f\|_\infty
}{
	\sqrt{M}
},
\label{eqFromAssumptions}
\end{equation}
where $c_{t-1}, \bar c_{t-1} <\infty$ are constants independent of $N$, $M$ and $\theta'$. 

To control the second term on the right hand side of \eqref{eqTriangle-Distort} we resort to assumption A.\ref{asPhiLipschitz}. In particular, note that for any $\theta',\theta'' \in D_\theta$ and any $f \in B(\Real^{d_x})$, we have
\begin{equation}
| (f,\phi_{t-1,\theta'}) - (f,\phi_{t-1,\theta''}) | \le b_{t-1}\|f\|_\infty \| \theta' - \theta'' \|
\label{eqPrepared-for-kernel}
\end{equation}
where the constant $b_{t-1} < \infty$ is independent of $\theta'$ and $\theta''$. Moreover, if $\theta''$ is random with probability distribution given by $\kappa_N^{\theta'}$, from assumption A.\ref{asKernel-2} we obtain that
\begin{eqnarray}
E\left[
	\left\|
		\theta' - \theta''
	\right\|^p
\right] &\le& \sup_{\theta' \in D_\theta} \int \left\|
	\theta' - \theta
\right\|^p \kappa_N^{\theta'}(d\theta) 
\le \frac{c_\kappa^p}{N^\frac{p}{2}}. \label{eqPintegrable}
\end{eqnarray}
Combining the inequalities \eqref{eqPrepared-for-kernel} and \eqref{eqPintegrable} yields
\begin{equation}
\| (f,\phi_{t-1,\theta'}) - (f,\phi_{t-1,\theta''}) \|_p \le \frac{ b_{t-1} c_\kappa \|f\|_\infty}{ \sqrt{N} }.
\label{eq2ndTerm}
\end{equation}

Finally, substituting \eqref{eq2ndTerm} and \eqref{eqFromAssumptions} into the triangle inequality \eqref{eqTriangle-Distort} completes the proof, with constants $\tilde c_{t-1}  = c_{t-1} + b_{t-1}c_\kappa$ and $\bar {\tilde c}_{t-1} = \bar c_{t-1}$.
$\QED$

Lemma \ref{lmRecursive} implies that we can ``leap'' from $\theta_{t-1}^{(i)}$ to $\bar \theta_t^{(i)}$ and still keep the associated particle filter in the inner layer running recursively, i.e., we do not have to start it over every time the particle position in the parameter space changes. If we incorporate some regularity assumptions on the likelihoods $g_{t,\theta}^{y_t}$, $t \ge 1$ (in such a way that we can resort to Lemma \ref{lmUnifConvPF}), then we arrive at an upper bound for the error $\| (h,\tilde \mu_t^{N,M}) - (h,\mu_t) \|_p$ after the weight update step. These assumptions are made explicit below.

\begin{Assumption} \label{asOnG}
Given a fixed sequence $Y_{1:T}=y_{1:T}$, the family of functions $\{ g_{t,\theta}^{y_t}; 1 \le t \le T, \theta \in D_\theta \}$ satisfies the following inequalities:
\begin{enumerate}

\item $\| g_t^{y_t} \|_\infty = \sup_{\theta \in D_\theta} \| g_{t,\theta}^{y_t} \|_\infty < \infty$ (which implies $\sup_{\theta \in D_\theta} u_t(\theta) = \sup_{\theta\in D_\theta} (g_{t,\theta}^{y_t},\xi_{t,\theta}) \le \| g_t^{y_t} \|_\infty$), and

\item $\inf_{\theta \in D_\theta} g_{t,\theta}^{y_t}(x)>0$ (which implies 
$
u_{t,\inf} = \inf_{\theta \in D_\theta} u_t(\theta) = \inf_{\theta \in D_\theta} (g_{t,\theta}^{y_t},\xi_{t,\theta}) > 0)
$
\end{enumerate}
for every $0 < t \le T$.
\end{Assumption}

\begin{Lema} \label{lmWeights-2}
Let $Y_{1:T}=y_{1:T}$ be fixed and choose any $0<t\le T$, any $h \in B(D_\theta)$ and any $f \in B(\Real^{d_x})$. Let $p \ge 1$ and assume that  A.\ref{asKernel-2}, A.\ref{asPhiLipschitz} and A.\ref{asOnG} hold. In Algorithm \ref{alRA}, if 
\begin{equation}
\| 
	(h,\bar \mu_{t-1}^{N,M}) - (h,\mu_{t-1})
\|_p \le \frac{c_{1,t}\|h\|_\infty}{\sqrt{N}} +  \frac{\bar c_{1,t}\|h\|_\infty}{\sqrt{M}}
\label{eqHypoLemmaWeight-2}
\end{equation}
for some constants $c_{1,t}, \bar c_{1,t} < \infty$ independent of $N$ and $M$, and the random measures $\{ \phi_{t-1,\theta_{t-1}^{(i)}}^M \}_{1 \le i \le N}$ satisfy
\begin{equation}
\sup_{1 \le i \le N} \|
	(f,\phi_{t-1,\theta_{t-1}^{(i)}}^M) - (f,\phi_{t-1,\theta_{t-1}^{(i)}}) 
\|_p \le \frac{
	k_{1,t-1}\|f\|_\infty
}{
	\sqrt{N}
} + \frac{
	\bar k_{1,t-1}\|f\|_\infty
}{
	\sqrt{M}
},
\label{eqOldPhis}
\end{equation}
for some constants $k_{1,t-1}, \bar k_{1,t-1} < \infty$ independent of $N$ and $M$, then
\begin{eqnarray}
\| 
	(h,\tilde \mu_{t}^{N,M}) - (h,\mu_{t}) 
\|_p &\le& \frac{c_{2,t}\|h\|_\infty}{\sqrt{N}} + \frac{\bar c_{2,t}\|h\|_\infty}{\sqrt{M}}, 
\label{eqBarMu} \\
\sup_{1 \le i \le N} \|
	(f,\xi_{t,\bar \theta_{t}^{(i)}}^M) - (f,\xi_{t,\bar \theta_{t}^{(i)}}) 
\|_p &\le& \frac{
	\tilde k_{2,t}\|f\|_\infty
}{
	\sqrt{N}
} + \frac{
	\bar {\tilde k}_{2,t}\|f\|_\infty
}{
	\sqrt{M}
}, \label{eqRecycledXi}\\
\sup_{1 \le i \le N} \|
	(f,\phi_{t,\theta_{t}^{(i)}}^M) - (f,\phi_{t,\theta_{t}^{(i)}}) 
\|_p &\le& \frac{
	k_{2,t}\|f\|_\infty
}{
	\sqrt{N}
} + \frac{
	\bar k_{2,t}\|f\|_\infty
}{
	\sqrt{M}
}
\label{eqRecycledPhi}
\end{eqnarray}
where the constants $c_{2,t}, \bar c_{2,t}, \tilde k_{2,t}, \bar {\tilde k}_{2,t}, k_{2,t}, \bar k_{2,t} < \infty$ are independent of $N$ and $M$. 
\end{Lema}

\noindent {\bf Proof:} Recall that the particle $\bar \theta_t^{(i)}$ is drawn from the kernel $\kappa_N^{\theta_{t-1}^{(i)}}(d\theta)$. Therefore, the inequality \eqref{eqOldPhis} together with Lemma \ref{lmRecursive} yields
\begin{equation}
\sup_{1\le i \le N} \| (f,\phi_{t-1,\theta_{t-1}^{(i)}}^M) - (f,\phi_{t-1,\bar \theta_t^{(i)}}) \|_p 
\le \frac{ \tilde c_{t-1}\|f\|_\infty }{ \sqrt{N} } + \frac{ \bar {\tilde c}_{t-1}\|f\|_\infty }{ \sqrt{M} }, 
\label{eqForL1-1}
\end{equation}
where the constants $\tilde c_{t-1}, \bar {\tilde c}_{t-1} < \infty$ are independent of $N$, $M$. However, the key feature of Algorithm \ref{alRA} is to set the approximation
\begin{equation}
\phi_{t-1,\bar \theta_t^{(i)}}^M \dfn \phi_{t-1,\theta_{t-1}^{(i)}}^M = \frac{1}{M} \sum_{j=1}^M \delta_{x_{t-1}^{(i,j)}}, \quad i=1, ..., N.
\nonumber 
\end{equation}
This choice of $\phi_{t-1,\bar \theta_t^{(i)}}^M$, together with the inequality \eqref{eqForL1-1} and Lemma \ref{lm1StepUnifConv}, yields the inequalities \eqref{eqRecycledXi} and \eqref{eqRecycledPhi} in the statement of Lemma \ref{lmWeights-2}.

Now we address the characterisation of the weights and, therefore, of the approximate measure $\tilde \mu_t^{N,M} = \sum_{i=1}^N w_t^{(i)} \delta_{\bar \theta_t^{(i)}}$. From the Bayes' theorem, the integral of $h$ w.r.t. $\mu_t$ can be written as
\begin{equation}
(h,\mu_t) = \frac{
	(u_t h, \mu_{t-1})
}{
	(u_t, \mu_{t-1})
}, \quad \mbox{while} \quad (h, \tilde \mu_t^{N,M}) = \frac{
	(u_t^M h, \bar \mu_{t-1}^{N,M})
}{
	(u_t^M, \bar \mu_{t-1}^{N,M})
}.
\end{equation}
Therefore, from the inequality \eqref{eqPreliminaries} we readily obtain
\begin{eqnarray}
| (h,\tilde \mu_t^{N,M}) - (h, \mu_{t-1}) | &\le& \frac{
	1
}{
	(u_t,\mu_{t-1})
} \left[
	\| h \|_\infty | (u_t^M, \bar \mu_{t-1}^{N,M}) - (u_t, \mu_{t-1}) |  
\right. \nonumber\\
&& \left.
	+ | (hu_t^M, \bar \mu_{t-1}^{N,M}) - (hu_t, \mu_{t-1}) |
\right],
\label{eqL1Error-0}
\end{eqnarray}
and \eqref{eqL1Error-0}, together with Minkowski's inequality, yields
\begin{eqnarray}
\| (h,\tilde \mu_t^{N,M}) - (h, \mu_{t-1}) \|_p &\le& \frac{
	1
}{
	(u_t,\mu_{t-1})
} \left[
	\| h \|_\infty \| (u_t^M, \bar \mu_{t-1}^{N,M}) - (u_t, \mu_{t-1}) \|_p 
\right. \nonumber\\
&& \left.
	+ \| (hu_t^M, \bar \mu_{t-1}^{N,M}) - (hu_t, \mu_{t-1}) \|_p,
\right]
\label{eqL1Error-1}
\end{eqnarray}
where $(u_t,\mu_{t-1})>0$ from assumption A.\ref{asOnG}-2

We need to find upper bounds for the two terms on the right hand side of \eqref{eqL1Error-1}. Consider first the term $\| (u_t^M,\bar \mu_{t-1}^{N,M}) - (u_t,\mu_{t-1}) \|_p$. A simple triangle inequality yields
\begin{equation}
\| (u_t^M,\bar \mu_{t-1}^{N,M}) - (u_t,\mu_{t-1}) \|_p \le 
\| (u_t^M,\bar \mu_{t-1}^{N,M}) - (u_t,\bar \mu_{t-1}^{N,M}) \|_p 
+ \| (u_t,\bar \mu_{t-1}^{N,M}) - (u_t,\mu_{t-1}) \|_p.
\label{eqWeights-triangle}
\end{equation}
On one hand, since $\sup_{\theta \in D_\theta} |u_t(\theta)| \le \| g_t^{y_t} \|_\infty < \infty$ (see A.\ref{asOnG}), it follows from the assumption in Eq. \eqref{eqHypoLemmaWeight-2} that
\begin{equation}
\| (u_t,\bar \mu_{t-1}^{N,M}) - (u_t,\mu_{t-1}) \|_p \le \frac{
	c_{1,t}\|g_t^{y_t}\|_\infty
}{
	\sqrt{N}
} + \frac{
	\bar c_{1,t}\|g_t^{y_t}\|_\infty
}{
	\sqrt{M}
}.
\label{eqBoundSecondTermWeight}
\end{equation}
On the other hand, we may note that
\begin{eqnarray}
| (u_t^M, \bar \mu_{t-1}^{N,M}) - (u_t,\bar \mu_{t-1}^{N,M}) |^p &=& \left|
	\frac{1}{N} \sum_{i=1}^N \left(
		u_t^M(\bar \theta_t^{(i)}) - u_t(\bar \theta_t^{(i)})
	\right)
\right|^p \nonumber \\
&\le& \frac{
	1
}{
	N
} \sum_{i=1}^N | u_t^M(\bar \theta_t^{(i)}) - u_t(\bar \theta_t^{(i)}) |^p,
\label{eqWeightsNearTheEnd}
\end{eqnarray}
which is readily obtained from Jensen's inequality.
However, the $i$-th term of the summation above is simply the ($p$-th power of the) approximation error of the integral $u_t(\bar \theta_t^{(i)}) = (g_{t,\bar \theta_t^{(i)}}^{y_t},\xi_{t,\bar \theta_t^{(i)}})$. Indeed, taking expectations on both sides of the inequality \eqref{eqWeightsNearTheEnd} yields
\begin{eqnarray}
E\left[
	\left|
		 (u_t^M, \bar \mu_{t-1}^{N,M}) - (u_t,\bar \mu_{t-1}^{N,M}) 
	\right|^p
\right] &\le& \frac{
	1
}{
	N
} \sum_{i=1}^N E\left[
	\left| 
		(g_{t,\bar \theta_t^{(i)}}^{y_t},\xi_{t,\bar \theta_t^{(i)}}^M) - (g_{t,\bar \theta_t^{(i)}}^{y_t},\xi_{t,\bar \theta_t^{(i)}}) 
	\right|^p
\right] \nonumber\\
&\le& \frac{1}{N} \sum_{i=1}^N \sup_{\theta \in D_\theta} \sup_{i \le 1 \le N} E\left[
	\left| 
		(g_{t,\theta}^{y_t},\xi_{t,\bar \theta_t^{(i)}}^M) - (g_{t,\theta}^{y_t},\xi_{t,\bar \theta_t^{(i)}}) 
	\right|^p
\right] \nonumber\\
\label{eqError_p}
\end{eqnarray}
From assumption A.\ref{asOnG} we have $\sup_{\theta \in D_\theta} \| g_{t,\theta}^{y_t} \|_\infty \le \| g_t^{y_t} \|_\infty$ and $\inf_{\theta \in D_\theta} g_{t,\theta}^{y_t}(x)>0$ for every $t=1, ..., T$ and every $x\in\Real^{d_x}$, hence Lemma \ref{lm1StepUnifConv} (see also Remark \ref{rmF_of_theta}) readily yields
\begin{equation}
\sup_{\theta \in D_\theta} \sup_{1 \le i \le N} E\left[
	\left| 
		(g_{t,\theta}^{y_t},\xi_{t,\bar \theta_t^{(i)}}^M) - (g_{t,\theta}^{y_t},\xi_{t,\bar \theta_t^{(i)}}) 
	\right|^p
\right] \le \frac{
	\hat k_{2,t}^p \| g_t^{y_t} \|_\infty^p
}{
	N^\frac{p}{2}
} + \frac{
	\bar {\hat k}_{2,t}^p \| g_t^{y_t} \|_\infty^p
}{
	M^\frac{p}{2}
}
\label{eqSupSup} 
\end{equation}
for some finite constants $\hat k_{2,t}$ and $\bar {\hat k}_{2,t}$ independent of $N$ and $M$. Substituting \eqref{eqSupSup} into \eqref{eqError_p} yields
\begin{equation}
E\left[
	\left|
		 (u_t^M, \bar \mu_{t-1}^{N,M}) - (u_t,\bar \mu_{t-1}^{N,M}) 
	\right|^p
\right]  \le \frac{
	\hat k_{2,t}^p \| g_t^{y_t} \|_\infty^p
}{
	N^\frac{p}{2}
} + \frac{
	\bar {\hat k}_{2,t}^p \| g_t^{y_t} \|_\infty^p
}{
	M^\frac{p}{2}
}
\nonumber
\end{equation}
or, equivalently, 
\begin{equation}
\| (u_t^M, \bar \mu_{t-1}^{N,M}) - (u_t,\bar \mu_{t-1}^{N,M}) \|_p \le \frac{
	\hat k_{2,t} \| g_t^{y_t} \|_\infty
}{
	\sqrt{N}
} + \frac{
	\bar {\tilde k}_{2,t} \| g_t^{y_t} \|_\infty
}{
	\sqrt{M}
}.
\label{eqBoundFirstTermWeight}
\end{equation}
Substituting \eqref{eqBoundFirstTermWeight} and \eqref{eqBoundSecondTermWeight} into \eqref{eqWeights-triangle} yields
\begin{equation}
\| (u_t^M, \bar \mu_{t-1}^{N,M}) - (u_t,\mu_{t-1}) \|_p \le \frac{
	c_t' \| g_t^{y_t} \|_\infty
}{
	\sqrt{N}
} + \frac{
	\bar c_t' \| g_t^{y_t} \|_\infty
}{
	\sqrt{M}
},
\label{eqBoundTerm_u}
\end{equation}
where $c_t' = c_{1,t} + \hat k_{2,t}$  and $\bar c_t' = \bar c_{1,t} + \bar {\hat k}_{2,t}$ are constants independent of $N$ and $M$.

Since $\| hu_t \|_\infty \le \| h \|_\infty \| g_t^{y_t} \|_\infty$ (the bound is independent of $\theta$), the same argument leading to the bound in \eqref{eqBoundTerm_u} can be repeated, step by step, on the norm $\| (hu_t^N,\bar \mu_{t-1}^N) - (hu_t,\mu_{t-1}) \|_p$, to arrive at
\begin{equation}
\| (hu_t^M, \bar \mu_{t-1}^{N,M}) - (hu_t,\mu_{t-1}) \|_p \le \frac{
	c_t'' \| h \|_\infty \|g_t^{y_t}\|_\infty
}{
	\sqrt{N}
} + \frac{
	\bar c_t'' \| h \|_\infty \|g_t^{y_t}\|_\infty
}{
	\sqrt{M}
},
\label{eqBoundTerm_hu}
\end{equation}
where $c_t'', \bar c_t''<\infty$ are constants independent of $N$ and $M$.

To complete the proof, we substitute \eqref{eqBoundTerm_u} and \eqref{eqBoundTerm_hu} back into \eqref{eqL1Error-1} and so obtain
\begin{equation}
\| (h,\tilde \mu_t^{N,M}) - (h, \mu_{t-1}) \|_p \le \frac{
	c_{2,t}\| h \|_\infty
}{
	\sqrt{N}
} + \frac{
	\bar c_{2,t}\| h \|_\infty
}{
	\sqrt{M}
}, \nonumber
\end{equation}
where the constants 
$
c_{2,t} = \| g_t^{y_t} \|_\infty \left( c_t' + c_t'' \right) / (u_t,\mu_{t-1}) < \infty
$ and
$
\bar c_{2,t} = \| g_t^{y_t} \|_\infty \left( \bar c_t' + \bar c_t'' \right) / (u_t,\mu_{t-1}) < \infty
$
are independent of $N$ and $M$. 
$\QED$

\subsection{Resampling}

We quantify the error in the resampling step 2(c) of Algorithm \ref{alRA}.

\begin{Lema} \label{lmResampling}
Let the sequence $Y_{1:T}=y_{1:T}$ be fixed and choose any $0 < t \le T$. If $h \in B(\Real^{d_\theta})$ and
\begin{equation}
\| 
	(h,\tilde \mu_{t}^{N,M}) - (h,\mu_{t})
\|_p \le \frac{c_{2,t}\| h \|_\infty}{\sqrt{N}} + \frac{\bar c_{2,t}\| h \|_\infty}{\sqrt{M}}
\label{eqAssResampling-theta}
\end{equation}
for some constants $c_{2,t}, \bar c_{2,t} < \infty$ independent of $N$ and $M$, then 
$$
\| 
	(h,\mu_{t}^{N,M}) - (h,\mu_{t}) 
\|_p \le \frac{c_{3,t}\| h \|_\infty}{\sqrt{N}} + \frac{\bar c_{3,t}\| h \|_\infty}{\sqrt{M}},
$$
where the constants $c_{3,t}, \bar c_{3,t} < \infty$ are independent of $N$ and $M$ as well.
\end{Lema}
\noindent {\bf Proof:} The proof of this Lemma is straightforward. The resampling step is the same as in a standard particle filter. See, e.g., the proof of \cite[Lemma 1]{Miguez13b} or simply the argument leading from Eq. \eqref{eqPpioResampling} to Eq. \eqref{eqFinalResampling} in Appendix \ref{ap1StepUnifConv}.
%
%
$\QED$

\subsection{Asymptotic convergence of the errors in $L_p$}

Finally, we can put Lemmas \ref{lmSampling}, \ref{lmWeights-2} and \ref{lmResampling} together in order to prove the convergence of the recursive Algorithm \ref{alRA}.

\begin{Teorema} \label{thConvergenceLp}
Let the sequence $Y_{1:T}=y_{1:T}$ be fixed ($T < \infty$), take an arbitrary test function $h\in B(\Real^{d_\theta})$, and assume that A.\ref{asKernel}--A.\ref{asOnG} 
hold. Then, for Algorithm \ref{alRA},
\begin{equation}
\|
	(h,\mu_{t}^{N,M}) - (h,\mu_{t}) 
\|_p \le \frac{c_{t}\|h\|_\infty}{\sqrt{N}} + \frac{\bar c_{t}\|h\|_\infty}{\sqrt{M}}, \quad 1 \le t \le T,
\label{eqConvergenceTheoremLp}
\end{equation}
where $\{ c_t, \bar c_t \}_{0 \le t \le T}$ is a sequence of constants independent of $N$ and $M$.
\end{Teorema}
\noindent {\bf Proof:} We prove \eqref{eqConvergenceTheoremLp} by induction in $t$. At time $t=0$, we draw $\theta_0^{(i)}$, $i=1, ..., N$, independently from the prior $\mu_0$ and it is straightforward to show that 
$
\|
	(h,\mu_{0}^{N,M}) - (h,\mu_{0}) 
\|_p \le \frac{c_{0} \| h \|_{\infty}}{\sqrt{N}}
$, where $c_0$ does not depend on $N$. Similarly, for each $i=1, ..., N$ we draw $M$ i.i.d. samples $\{ x_0^{(i,j)} \}_{1 \le j \le M}$ from the distribution with measure $\tau_0$ and it is not difficult to check that the random measures $\phi_{0,\theta_0^{(i)}}^M = \frac{1}{M}\sum_{j=1}^M \delta_{x_0^{(i,j)}}$ satisfy 
$$
\| (f,\phi_{0,\theta_0^{(i)}}^M) - (f,\phi_{0,\theta_0^{(i)}}) \|_1 \le \frac{ \bar k_0 \| f \|_\infty }{ \sqrt{M} }
$$  
for every $i \in \{ 1, ..., N \}$ and any $f \in B(\Real^{d_x})$. The constant $k_0$ is independent of $M$ and $\{ \theta_0^{(i)} \}_{1 \le i \le N}$ (note that $\tau_0 = \phi_{0,\theta}$ is actually independent of $\theta$). 

Assume that, at time $t-1$,
$$
\|
	(h,\mu_{t-1}^{N,M}) - (h,\mu_{t-1}) 
\|_p \le \frac{c_{t-1}\|h\|_\infty}{\sqrt{N}} + \frac{\bar c_{t-1}\|h\|_\infty}{\sqrt{M}},
$$
where $c_{t-1}, \bar c_{t-1} < \infty$ are independent of $N$ and $M$, and, for any $f\in B(\Real^{d_x})$,
\begin{equation}
\sup_{1 \le i \le N} \| (f,\phi_{t-1,\theta_{t-1}^{(i)}}^M) - (f,\phi_{t-1,\theta_{t-1}^{(i)}}) \|_p \le 
\frac{ k_{t-1}\|f\|_\infty }{ \sqrt{N} } + \frac{ \bar k_{t-1}\|f\|_\infty }{ \sqrt{M} },
\nonumber
\end{equation}
where $k_{t-1}, \bar k_{t-1} < \infty$ are constants independent of $N$ and $M$. Then, we simply ``concatenate'' Lemmas \ref{lmSampling}, \ref{lmWeights-2} and \ref{lmResampling} (in that order) to obtain
\begin{eqnarray}
\|
	(h,\mu_t^{N,M}) - (h,\mu_t) 
\|_p &\le& \frac{c_{t}\|h\|_\infty}{\sqrt{N}} + \frac{\bar c_{t}\|h\|_\infty}{\sqrt{M}}, \nonumber\\
\sup_{1\le i \le N} \| (f,\phi_{t,\theta_{t}^{(i)}}^M) - (f,\phi_{t,\theta_{t}^{(i)}}) \|_p &\le& \frac{ k_{t}\| f \|_\infty }{ \sqrt{N} } + \frac{ \bar k_{t}\| f \|_\infty }{ \sqrt{M} }, \label{eqConvergenceConditionalFilters}
\end{eqnarray}
for some constants $c_t, \bar c_t, k_t, \bar k_t < \infty$ independent of $N$ and $M$. $\QED$

\begin{Nota} \label{rmConvergenceInnerFilters}
The argument of the proof of Theorem \ref{thConvergenceLp} also yields, as a by-product, error rates for the (approximate) conditional filters $\phi_{t,\theta_t^{(i)}}^M$ computed for each particle in the parameter space, as shown by the inequality in \eqref{eqConvergenceConditionalFilters}. These rates are uniform for any $\theta \in D_\theta$.
%
%
%
\end{Nota}


\subsection{Approximation of the joint measure $\pi_t$} \label{ssJoint}

Integrals w.r.t. the joint measure $\pi_t$ introduced in \eqref{eqJointDef} can be written naturally in terms of the marginal measures $\phi_{t,\theta}$ and $\mu_t$. To be specific, choose any integrable function $f:D_\theta \times \Real^{d_x} \rw \Real$ and define $f^\theta:\Real^{d_x}\rw\Real$, where $f^\theta(x_t) \dfn f(\theta,x_t)$, and ${\sf f}_t: D_\theta \rw \Real$, where ${\sf f}_t(\theta) \dfn \int f^\theta(x_t) \phi_{t,\theta}(dx_t) = (f^\theta,\phi_{t,\theta})$. Then we can write
\begin{equation}
(f,\pi_t) = \int \int f(\theta,x_t)\pi_t(d\theta,dx_t) = \int {\sf f}_t(\theta) \mu_t(d\theta) = ({\sf f}_t, \mu_t).
\label{eqIntegJoint-1}
\end{equation}
It is straightforward to approximate $\pi_t$ as
$$
\pi_t^{N,M}(d\theta \times dx_t) = \frac{1}{NM}\sum_{i=1}^N\sum_{j=1}^M \delta_{\theta_t^{(i)},x_t^{(i,j)}}(d\theta \times dx_t),
$$
which yields 
\begin{equation}
(f,\pi_t^{N,M}) = \frac{1}{NM}\sum_{i=1}^N\sum_{j=1}^M f(\theta_t^{(i)},x_t^{(i,j)}) = ({\sf f}_t^M, \mu_t^N),
\label{eqIntegJointApprox}
\end{equation}
where ${\sf f}_t^M(\theta_t^{(i)}) = (f^{\theta^{(i)}_t},\phi_{t,\theta_t^{(i)}}^M)$.

It is relatively easy to use the results obtained earlier in this Section in order to show that, for any $f \in B(D_\theta \times \Real^{d_x})$, the $L_p$ error norm $\| (f,\pi_t^{N,M}) - (f,\pi_t) \|_p$ has an upper bound of order $\frac{1}{\sqrt{N}} + \frac{1}{\sqrt{M}}$.
\begin{Teorema} \label{thConvergenceL1-Joint}
Let the sequence $Y_{1:T}=y_{1:T}$ be fixed, take an arbitrary test function $f \in B(D_\theta \times \Real^{d_\theta})$ and assume that A.\ref{asKernel}--A.\ref{asOnG}
hold. Then, for any $p \ge 1$, Algorithm \ref{alRA} yields
\begin{equation}
\|
	(f,\pi_{t}^{N,M}) - (f,\pi_{t}) 
\|_p \le \frac{c_{t}\| f \|_\infty}{\sqrt{N}} + \frac{\bar c_{t}\| f \|_\infty}{\sqrt{M}}, \quad 1 \le t \le T,
\label{eqConvergenceTheoremL1-Joint}
\end{equation}
where $\{ c_t,\bar c_t \}_{1 \le t \le T}$ is a sequence of finite constants independent of $N$ and $M$.
\end{Teorema}
\noindent {\bf Proof:} From Eqs. \eqref{eqIntegJoint-1} and \eqref{eqIntegJointApprox}, $(f,\pi_t^{N,M})-(f,\pi_t)=({\sf f}_t^M,\mu_t^{N,M}) - ({\sf f}_t,\mu_t)$ and a triangle inequality yields
\begin{equation}
\|
	({\sf f}_t^M,\mu_{t}^{N,M}) - ({\sf f}_t,\mu_{t}) 
\|_p \le \| 
	({\sf f}_t^M,\mu_t^{N,M}) - ({\sf f}_t,\mu_t^{N,M}) 
\|_p + \| 
	({\sf f}_t,\mu_t^{N,M})  - ({\sf f}_t,\mu_{t}) 
\|_p.
\label{eqTriangleJoint}
\end{equation}
Since ${\sf f_t} \in B(D_\theta)$ (namely, $\| {\sf f}_t \|_\infty \le \| f \|_\infty$), Theorem \ref{thConvergenceLp} yields a bound for the second term on the right hand side of \eqref{eqTriangleJoint}, i.e.,
\begin{equation}
\| 
	({\sf f}_t,\mu_t^{N,M})  - ({\sf f}_t,\mu_{t}) 
\|_p \le \frac{
	\hat c_t \| f \|_\infty
}{
	\sqrt{N}
} + \frac{
	\bar {\hat c}_t \| f \|_\infty
}{
	\sqrt{M}
},
\label{eqSecondTermJoint}
\end{equation}
where $\hat c_t, \bar {\hat c}_t < \infty$ are constants independent of $N$ and $M$. 

In order to control the first term on the right hand side of \eqref{eqTriangleJoint}, we note that
\begin{eqnarray}
E\left[
	\left|
		({\sf f}_t^M,\mu_t^{N,M}) - ({\sf f}_t,\mu_t^{N,M}) 
	\right|^p 
\right] &\le& \frac{1}{N} 
	\sum_{i=1}^N E\left[
		\left|
			(f^{\theta_t^{(i)}},\phi_{t,\theta_t^{(i)}}^M) - (f^{\theta_t^{(i)}},\phi_{t,\theta_t^{(i)}})
		\right|^p
	\right]
\label{eqItsnotequal} \\
&\le& \sup_{\theta \in D_\theta} \sup_{1 \le i \le N}  E\left[
	\left|
		 (f^\theta,\phi_{t,\theta_t^{(i)}}^M) - (f^\theta,\phi_{t,\theta_t^{(i)}})
	\right|^p
\right],
\nonumber
\end{eqnarray}
where \eqref{eqItsnotequal} follows from Jensen's inequality. 
However, since $f^\theta \le \| f \|_\infty < \infty$, we can resort to Remark \ref{rmConvergenceInnerFilters} in order to obtain 
$$
\sup_{1 \le i \le N} E\left[
	\left|
		 (f^\theta,\phi_{t,\theta_t^{(i)}}^N) - (f^\theta,\phi_{t,\theta_t^{(i)}})
	\right|^p
\right] \le \frac{
	k_t^p \| f \|_\infty^p
}{
	N^\frac{p}{2}
} + \frac{
	\bar k_t^p \| f \|_\infty^p
}{
	M^\frac{p}{2}
},
$$
where the constants $k_t, \bar k_t<\infty$ are independent of $N$ and $M$. Since the latter upper bound is uniform over $D_\theta$ (recall Remark \ref{rmF_of_theta}), it follows that 
\begin{eqnarray}
E\left[
	\left|
		({\sf f}_t^M,\mu_t^{N,M}) - ({\sf f}_t,\mu_t^{N,M}) 
	\right|^p 
\right] &\le& \sup_{\theta \in D_\theta} \sup_{1 \le i \le N}  E\left[
	\left|
		 (f^\theta,\phi_{t,\theta_t^{(i)}}^N) - (f^\theta,\phi_{t,\theta_t^{(i)}})
	\right|^p
\right] \nonumber\\
&\le& \frac{
	k_t^p \| f \|_\infty^p
}{
	N^\frac{p}{2}
} + \frac{
	\bar k_t^p \| f \|_\infty^p
}{
	M^\frac{p}{2}
} \nonumber
\end{eqnarray}
as well or, equivalently,
\begin{equation}
\| ({\sf f}_t^M,\mu_t^{N,M}) - ({\sf f}_t,\mu_t^{N,M}) \|_p \le \frac{k_t\|f\|_\infty}{\sqrt{N}} + \frac{\bar k_t\|f\|_\infty}{\sqrt{M}}.
\label{eqFirstTermJoint}
\end{equation}

Substituting \eqref{eqFirstTermJoint} and \eqref{eqSecondTermJoint} into the triangle inequality \eqref{eqTriangleJoint} yields the desired result, with constants $c_t = \hat c_t + k_t$ and $\bar c_t = \bar {\hat c}_t + \bar k_t$, $1 \le t \le T$, independent of $N$ and $M$. 
$\QED$

\subsection{Effective sample size} \label{ssESS}

After completing all operations at time $t-1$, Algorithm \ref{alRA} produces a system of particles $\{ \theta_{t-1}^{(i)} \}_{1 \le i \le N}$, where many of its elements may be located at the same position in the parameter space because of the resampling step. At time $t$, the first operation of Algorithm \ref{alRA} is the jittering of the particles in order to restore their diversity. After jittering, the new system $\{ \bar \theta_t^{(i)} \}_{1 \le i \le N}$ is available. However, depending on the choice of kernel $\kappa_N$, it is possible that {\em not} every particle in $\{ \theta_{t-1}^{(i)} \}_{1 \le i \le N }$  has actually been changed, hence the jittered system $\{ \bar \theta_t^{(i)} \}_{1 \le i \le N}$ may still contain replicated elements, i.e., particles with different indices that correspond to the same position in the parameter space $D_\theta$.

Let $\hat N_t$ denote the number of distinct particles in the system $\{ \bar \theta_t^{(i)} \}_{1 \le i \le N}$ and let $\{ \tilde \theta_t^{(i)} \}_{1 \le i \le \hat N_t}$ be the set of those distinct particles. Obviously, $1 \le \hat N_t \le N$. We use $n_t^{(i)}$ to denote the number of replicas of $\tilde \theta_t^{(i)}$ included in the original system $\{ \bar \theta_t^{(i)} \}_{ 1 \le i \le N}$. It is straightforward to check that, for every $i=1, ..., \hat N_t$,
\begin{equation}
1 \le n_t^{(i)} \le N - \hat N_t + 1,
\nonumber
\end{equation}
while 
$
\sum_{i=1}^{\hat N_t} n_t^{(i)} = N.
$

The size of the set $\{ \tilde \theta_t^{(i)} \}_{1 \le i \le \hat N_t}$ is particularly relevant to the computation of the so-called effective sample size (ESS) \citep{Kong94} (see also \citep{Doucet00}) of the particle approximation produced by Algorithm \ref{alRA}. The ESS, which is commonly used to assess the numerical stability of particle filters \citep{Chopin12,Beskos14}, was defined in \citep{Kong94} as
\begin{equation}
\mbox{ESS}_t(N) = \frac{
	N
}{
	1 + V_t^2
},
\nonumber
\end{equation}
where $V_t^2$ denotes the variance of the non-normalised importance weights (namely, the variance of $u_t^M(\theta)$ in the case of Algorithm \ref{alRA}). Since this variance cannot be computed in closed form, the ESS has to be estimated and the most commonly used estimator takes the form \citep{Kong94,Doucet00}
\begin{eqnarray} 
\widehat{\mbox{ESS}}_t(N) &=& \frac{
	1
}{
	\sum_{i=1}^N w_t^{(i)^2}
} \nonumber \\ 
&=& \frac{
	\left(
		\sum_{i=1}^N u_t^M(\bar \theta_t^{(i)})
	\right)^2
}{
	\sum_{i=1}^N u_t^M(\bar \theta_t^{(i)})^2
} \label{eqk0} \\
&=& \frac{
	\left(
		\sum_{i=1}^{\hat N_t} n_t^{(i)} u_t^M(\tilde \theta_t^{(i)})
	\right)^2
}{
	\sum_{i=1}^{\hat N_t} n_t^{(i)} u_t^M(\tilde \theta_t^{(i)})^2
},
\label{eqk1}
\end{eqnarray}
where \eqref{eqk0} follows from the construction of the normalised weights in Algorithm \ref{alRA} and in \eqref{eqk1} we write the estimator explicitly in terms of the system of distinct particles\footnote{We assume that the algorithm is implemented efficiently, meaning that when a subset of particles is found to correspond to the same position in the parameter space the likelihood of that position is estimated only once. In other words, if we have indices $i_0, i_1, \ldots, i_{n_t^{(i_0)}}$ such that $\tilde \theta_t^{(i_0)} = \bar \theta_t^{(i_1)} = \ldots = \bar \theta_t^{\left(i_{n_t^{(i_o)}}\right)}$, then we compute $u_t^M(\tilde \theta_t^{(i_0)})$ only once.} $\{ \tilde \theta_t^{(i)} \}_{1 \le i \le \hat N_t}$.

The estimator of the ESS in Eq. \eqref{eqk1} takes values between 1 and $N$, with 1 being the worst and $N$ being the best outcome. However, it can become uninformative when we actually have replicated particles, i.e., when $\hat N_t < N$. To see the problem, let us consider the extreme case in which $\hat N_t=1$ and, as a consequence, $n_t^{(1)} = N$. If we substitute these values in \eqref{eqk1} and realise that $\sum_{i=1}^{\hat N_t} n_t^{(i)} u_t^M(\tilde \theta_t^{(i)}) = N u_t^M(\tilde \theta_t^{(1)})$, then we readily obtain that $\widehat{\mbox{ESS}}_t(N)=N$. This seems to indicate that we have an ``optimal'' set of particles, as the maximum ESS is attained, when it is actually a fully degenerate set with one single particle replicated $N$ times. This difficulty does not arise in standard particle filtering applications because the ESS is typically estimated after the weight update step, before resampling, when all particles are different with probability 1. 

To overcome this problem, we propose to use a different estimator of the ESS. Recall that $w_t^{(i)} = \frac{u_t^M(\bar \theta_t^{(i)})}{\sum_{k=1}^N u_t^M(\bar \theta_t^{(k)} }$, $1 \le i \le N$, are the normalised weights. When there are multiple samples at the same position in $D_\theta$, the resulting probability measure
\begin{equation}
\mu_t^{N,M} = \sum_{i=1}^N w_t^{(i)} \delta_{\bar \theta_t^{(i)}}
\nonumber
\end{equation}
can be rewritten as
\begin{equation}
\mu_t^{N,M}= \sum_{i=1}^{\hat N_t} v_t^{(i)} \delta_{\tilde \theta_t^{(i)}},
\label{kk13}
\end{equation}
where $v_t^{(i)} = n_t^{(i)}w_t^{(i)}$ is the probability mass that $\mu_t^{N,M}$ allocates at position $\tilde \theta_t^{(i)}$. If we are given $\mu_t^{N,M}$ in the form of \eqref{kk13}, a fairly natural estimator the ESS is
\begin{equation}
\overline{\mbox{ESS}}_t(N) = \frac{
	1
}{
	\sum_{i=1}^{\hat N_t} \left( v_t^{(i)} \right)^2
} = \frac{
	\left(
		\sum_{k=1}^N u_t^M(\bar \theta_t^{(k)})
	\right)^2
}{
	\sum_{i=}^{\hat N_t} \left( 
		n_t^{(i)} u_t^M(\tilde \theta_t^{(i)})
	\right)^2
}
\label{kk14}
\end{equation}  
where we note that $\sum_{k=1}^{\hat N_t} n_t^{(i)} u_t(\tilde \theta_t^{(k)}) = \sum_{k=1}^N u_t(\bar \theta_t^{(k)})$.

When all the particles are distinct, $\hat N_t = N$ and $n_t^{(i)} = 1$ for every $i$, the estimator in \eqref{kk14} reduces to the standard one in \eqref{eqk1}. On the other hand, when $\hat N_t = 1$ and $n_t^{(1)} = N$, the formula in \eqref{kk14} yields $\overline{\mbox{ESS}}_t(N) = 1$, which is the minimal ESS and the expected result in this fully degenerate case. We recall that $\widehat{\mbox{ESS}}_t(N)=N$ in the same scenario. 
Finally, if we divide the expression in \eqref{kk14} by $N$ then we obtain an estimate of the normalised ESS (NESS) \citep{Doucet00} of the form 
\begin{equation} 
\overline{\mbox{NESS}}_t(N) =\frac{
	\left(
		\sum_{k=1}^N u_t^M(\bar \theta_t^{(k)})
	\right)^2
}{
	N \sum_{i=1}^{\hat N_t} \left( 
		n_t^{(i)} u_t^M(\tilde \theta_t^{(i)})
	\right)^2
}
\label{eqk2}
\end{equation}
that takes values in the interval $[ N^{-1}, 1 ]$. 

\section{A numerical example} \label{sExamples}

Let us consider the problem of jointly tracking the dynamic variables and estimating the fixed parameters of a 3-dimensional Lorenz system \citep{Lorenz63} with additive dynamical noise and partial observations \citep{Chorin04}. To be specific, consider a 3-dimensional stochastic process $\{ X(s) \}_{s\in(0,\infty)}$ taking values on $\Real^3$, whose dynamics is described by the system of stochastic differential equations
$$
dX_1 = -S(X_1-Y_1) + dW_1, \quad 
dX_2 = RX_1 - X_2 - X_1X_3 + dW_2, \quad 
dX_3 = X_1X_2 - BX_3 + dW_3, \nonumber
$$
where $\{ W_i(s) \}_{s\in(0,\infty)}$, $i=1, 2, 3$, are independent 1-dimensional Wiener processes and $(S,R,B)\in\Real$ are static model parameters. A discrete-time version of the latter system using the Euler-Maruyama method with integration step $\intstep>0$ is straightforward to obtain and yields the model
\begin{eqnarray}
X_{1,t} &=& X_{1,t-1} - \intstep S(X_{1,t-1}-X_{2,t-1}) + \sqrt{\intstep} U_{1,t},\label{eqDiscreteLorenz-1}\\
X_{2,t} &=& X_{2,t-1} + \intstep ( RX_{1,t-1} - X_{2,t-1} - X_{1,t-1}X_{3,t-1} ) + \sqrt{\intstep} U_{2,t}, \label{eqDiscreteLorenz-2}\\
X_{3,t} &=& X_{3,t-1} + \intstep ( X_{1,t-1}X_{2,t-1} - BX_{3,t-1} ) + \sqrt{\intstep} U_{3,t}, \label{eqDiscreteLorenz-3}
\end{eqnarray} 
where $\{ U_{i,t} \}_{t=0, 1, ...}$, $i=1,2,3$, are independent sequences of i.i.d. normal random variables with 0 mean and variance 1. System \eqref{eqDiscreteLorenz-1}-\eqref{eqDiscreteLorenz-3} is partially observed every 40 discrete-time steps, i.e., we collect a sequence of 2-dimensional observations $\{ Y_n=(Y_{1,n},Y_{3,n}) \}_{n=1, 2, ...}$, of the form 
\begin{equation}
Y_{1,n} = k_o X_{1,40n} + V_{1,n}, \quad
Y_{3,n} = k_o X_{3,40n} + V_{3,n}, \label{eqObservLorenz}
\end{equation}
where $k_o >0$ is a fixed scale parameter and $\{ V_{i,n} \}_{n=1, 2, ...}$, $i=1,3$, are independent sequences of i.i.d. normal random variables with zero mean and variance $\sigma^2 = \frac{1}{10}$.

Let $X_t=(X_{1,t},X_{2,t},X_{3,t})$ be the state vector, let $Y_n=(Y_{1,n},Y_{3,n})$ be the observation vector and let $\Theta = (S,R,B,k_o)$ be the set of model parameters to be estimated. The dynamic model given by Eqs. \eqref{eqDiscreteLorenz-1}--\eqref{eqDiscreteLorenz-3} yields the family of kernels $\tau_{t,\theta}(dx|x_{t-1})$ and the observation model of Eq. \eqref{eqObservLorenz} yields the likelihood function $g_{n,\theta}^{y_n}(x_n)$, both in a straightforward manner. The goal is to track the sequence of joint posterior probability measures $\pi_n$, $n=1, 2, ...$, for $\{ \hat X_n, \Theta \}_{n=1, ...}$, where $\hat X_n = X_{40n}$. Note that one can draw a sample $\hat X_n = \hat x_n$ conditional on some $\theta$ and $\hat X_{n-1} = \hat x_{n-1}$ by successively simulating
$$
\tilde x_t \sim \tau_{t,\theta}(dx|\tilde x_{t-1}), \quad t=40(n-1)+1, ..., 40n, 
$$
where $\tilde x_{40(n-1)} = \hat x_{n-1}$ and $\hat x_n = \tilde x_{40n}$. For the sake of the example, the prior probability measure for the parameters, $\mu_0(d\theta)$, is chosen to be uniform, namely
$$
S \sim \mU(5, 20), \quad R \sim \mU(18, 50), \quad R \sim \mU(1,8) \quad \mbox{and} \quad k_o \in \mU(0.5, 3),
$$ 
where $\mU(a,b)$ is the uniform probability distribution in the interval $(a,b)$. The prior measure for the state variables is normal, namely
$
X_0 \sim \mN(x_*,v_0^2 \mI_3),
$
where $x_* = (-5.91652; -5.52332; 24.5723)$ is the mean and $v_0^2\mI_3$ is the covariance matrix, with $v_0^2 = 10$. (The value $x_*$ is taken from a typical run of the deterministic Lorenz 63 model, once in its stationary regime.)

We have applied the nested particle filter (Algorithm \ref{alRA}), with $N=M$ (i.e., the same number of particles in the outer and inner filters, following Remark \ref{rmMuPi}), to estimate the fixed parameters $S, R, B$ and $k_o$. Besides selecting the total number of particles $K=NM$, the only ``tuning'' necessary for the algorithm is the choice of the jittering kernel. One of the simplest possible choices is to jitter each parameter independently from the others, using Gaussian distributions truncated to fit the support of each parameter. To be specific, let $\mTN(\mu,\sigma^2,A,B)$ denote the Gaussian distribution with mean $\mu$ and variance $\sigma^2$ truncated to have support on the interval $(A,B)$, i.e., the distribution with pdf
$$
p_\mTN (x;\mu,\sigma^2,A,B) = \frac{
	\exp\left\{ 
		\frac{1}{2\sigma^2} (x-\mu)^2 
	\right\}
}{
	\int_A^B \exp\left\{ 
		\frac{1}{2\sigma^2} (z-\mu)^2
	\right\} dz
}.
$$
We choose the jittering kernel $\kappa_N^{\theta'}$, with $\theta' = (S',R',B',k_o')$, to be 
the conditional probability distribution with density
\begin{eqnarray}
\kappa_N^{S',R',B',k_o'}(S,R,B,k_o) &=& p_\mTN(S;S',\sigma_{N,S}^2,5,20) \times
p_\mTN(R;R',\sigma_{N,R}^2,18,50) \nonumber \\
&&\times
p_\mTN(B;B',\sigma_{N,B}^2,1,8) \times
p_\mTN(k_o;k_o',\sigma_{N,k_o}^2,0.5,3). \nonumber
\end{eqnarray}
This choice of kernel is possibly far from optimal (in terms or estimation accuracy) but it is simple and enables us to show that Algorithm \ref{alRA} works without having to fit a sophisticated kernel.
 
If we are merely interested in estimating the parameter values, then the test function $h \in B(D_\theta)$ in Theorem \ref{thConvergenceLp} is simply the projection of the parameter vector on the desired component, i.e., for $\theta=(\theta_1, ..., \theta_4) = (S,R,B,k_o)$ we are interested in the functions $h_i(\theta) = \theta_i, \quad i=1, ..., 4$. Therefore, the estimator of the parameter $\theta_i$ at time $t$ has the form
$$
\theta_{i,t}^{N,N} = (h_i,\mu_t^{N,N}) = \frac{1}{N}\sum_{j=1}^N h_i(\theta_t^{(j)}), \quad i=1, ..., 4.
$$
Furthermore, if we aim at the minimising the $L_1$ errors, $E\left[ | \theta_{i,t}^N - \theta_i | \right]$, Proposition \ref{propPeque} in Appendix \ref{apPequeLema} shows that it is enough to choose the jittering variances as 
$$
(\sigma_{N,S}^2,\sigma_{N,R}^2,\sigma_{N,B}^2,\sigma_{N,k_o}^2) = \frac{1}{N^\frac{3}{2}} (c_S,c_R,c_B,c_{k_o})
$$
for arbitrary positive constants $c_S, c_R, c_B$ and  $c_{k_o}$ in order to satisfy the assumptions A.\ref{asKernel} and A.\ref{asKernel-2}. For the simulations in this section we have set $(c_S,c_R,c_B,c_{k_o}) = (60,60,10,1)$ (we roughly choose bigger constants for the parameters with bigger support).

\begin{figure}
\centerline{
        \subfigure[Parameter $S$.]{
                \includegraphics[width=0.35\linewidth]{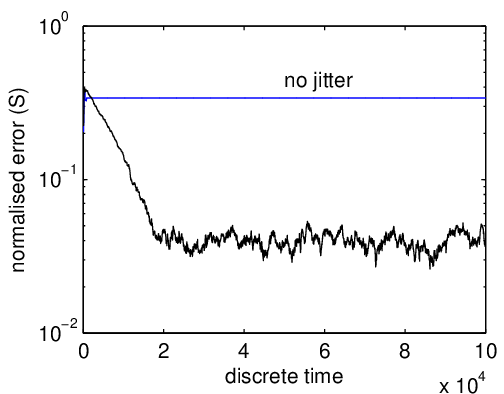}
        }
        \subfigure[Parameter $R$.]{
                \includegraphics[width=0.35\linewidth]{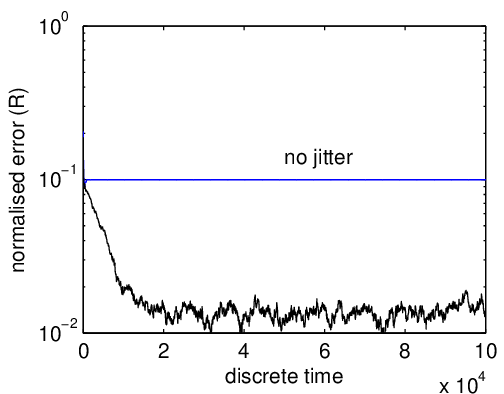}
        }
}
\centerline{
        \subfigure[Parameter $B$.]{
                \includegraphics[width=0.35\linewidth]{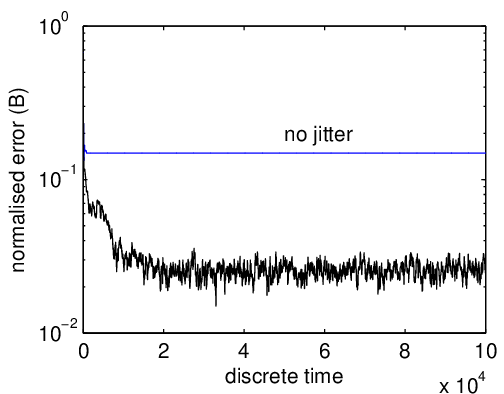}
        }
        \subfigure[Parameter $k_o$.]{
                \includegraphics[width=0.35\linewidth]{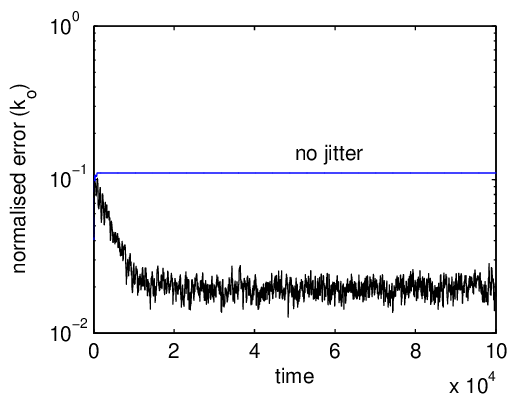}
        }
}
\caption{Average of the absolute parameter estimation errors over 50 independent simulation runs using Algorithm \ref{alRA} with $N=M=300$ particles ($K=N^2=90 \times 10^3$ particles overall). The absolute errors are normalised w.r.t. the true parameter values, $S=10, R=28, B=\frac{8}{3}$ and $k_o=\frac{4}{5}$. The results obtained when jittering is suppressed in Algorithm \ref{alRA} (labeled as {\em no jitter}) are shown for comparison. The horizontal axis is in discrete-time units. As the integration period is $\intstep=10^{-3}$, $100,000$ discrete-time steps amount 100 continuous time units. Observations are collected every 40 discrete-time steps.}
\label{fMeanError_N300_t40}
\end{figure}

Figure \ref{fMeanError_N300_t40} shows the average, over 50 independent simulations, of the normalised absolute errors $|\theta_{i,t}^{N,N}-\theta_i|/\theta_i$ versus continuous time when we run Algorithm \ref{alRA} with $N=M=300$. The figure shows how the errors converge over time (as $\mu_t$ concentrates around the true value $\theta = (10,28,8/3,0.8)$). We have also included the errors attained by a modified version of Algorithm \ref{alRA} in which the jittering step is removed. It is seen that the particle representation of $\mu_t$ soon collapses and the algorithm {\em without} jittering turns out unable to estimate the parameters. The integration period for all the simulations shown in this section is $\intstep=10^{-3}$, hence $100 \times 10^3$ discrete-time steps amount to 100 continuous time units. Observations are collected every 40 discrete steps. Even for this relatively simple system, running a non-recursive algorithm such as SMC$^2$ becomes impractical (recall that the computational complexity of the SMC$^2$ method increases quadratically with the number of discrete-time steps).

\begin{figure}
\centerline{
        \subfigure[Parameter $S$. The least squares fit of the errors yields $c \approx 0.807$.]{
                \includegraphics[width=0.35\linewidth]{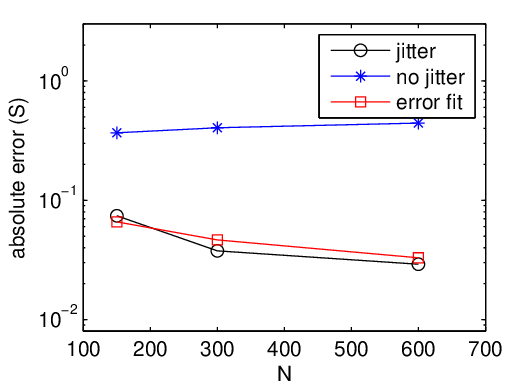}
        }
        \subfigure[Parameter $R$. The least squares fit of the errors yields $c \approx 0.290$.]{
                \includegraphics[width=0.35\linewidth]{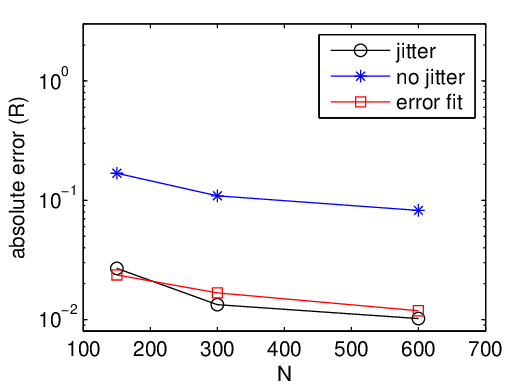}
        }
}
\centerline{
        \subfigure[Parameter $B$. The least squares fit of the errors yields $c\approx 0.496$.]{
                \includegraphics[width=0.35\linewidth]{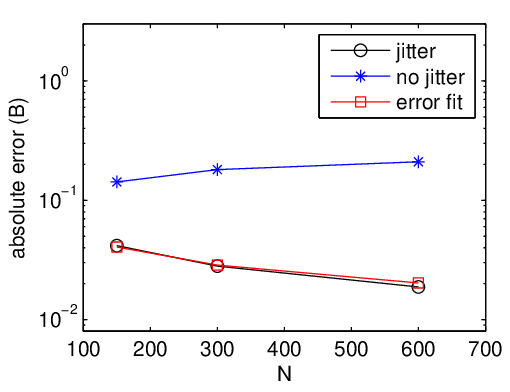}
        }
        \subfigure[Parameter $k_o$. The least squares fit of the errors yields $c\approx 0.397$.]{
                \includegraphics[width=0.35\linewidth]{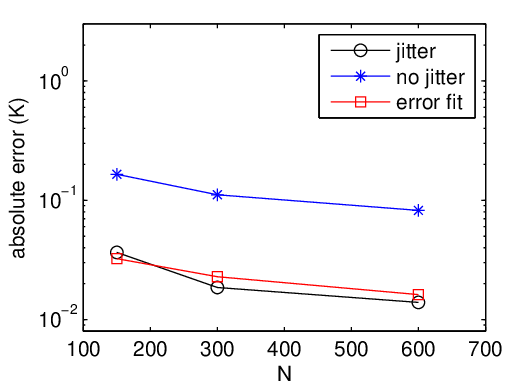}
        }
}
\caption{Average of the absolute parameter estimation errors over 20 independent simulation runs using Algorithm \ref{alRA} with $N=M=150$, $N=M=300$ and $N=M=600$ (the total number of particles is $N^2$). The errors are normalised w.r.t. the true parameter values, $S=10, R=28, B=\frac{8}{3}$ and $k_o=\frac{4}{5}$. The curves labeled {\em error fit} have the form $\frac{c}{\sqrt{N}}$, where the constant $c$ is a least squares estimate computed independently for each parameter. The results obtained when jittering is suppressed in Algorithm \ref{alRA} (labeled as {\em no jitter}) are also shown for comparison. In each simulation, the Lorenz system was run for 24,000 discrete-time steps ($24$ continuous-time steps, for $\intstep=10^{-3}$), with observations collected every $40$ discrete steps.}
\label{fMeanError_vs_N_t24}
\end{figure}

In Figure \ref{fMeanError_vs_N_t24} we plot the average of the normalised errors versus the number of particles in Algorithm \ref{alRA} (namely, for $N=150, 300, 600$). We have carried out 20 independent simulation trials (per point in the plot). In each simulation, the Lorenz system is run from continous time 0 to 24 (i.e., $24,000$ discrete time steps), with the errors computed by averaging $|\theta_{i,t}^{N,N}-\theta_i|/\theta_i$ over the continuous time interval (22,24). As in Figure \ref{fMeanError_N300_t40}, the performace of Algorithm \ref{alRA} with the jittering step removed is also displayed, and again we observe how it fails to yield accurate parameter estimates. For the outputs of Algorithm \ref{alRA} {\em with} jittering, we also display a least squares fit of the function $e(N) = \frac{c}{\sqrt{N}}$ to the averaged errors (with $c$ constant w.r.t. $N$), as suggested by Theorem \ref{thConvergenceLp}. 

\begin{figure}
\centerline{
        \subfigure[Parameter $S$.]{
                \includegraphics[width=0.35\linewidth]{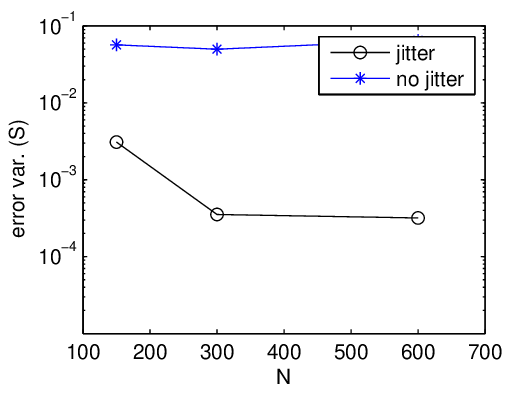}
        }
        \subfigure[Parameter $R$.]{
                \includegraphics[width=0.35\linewidth]{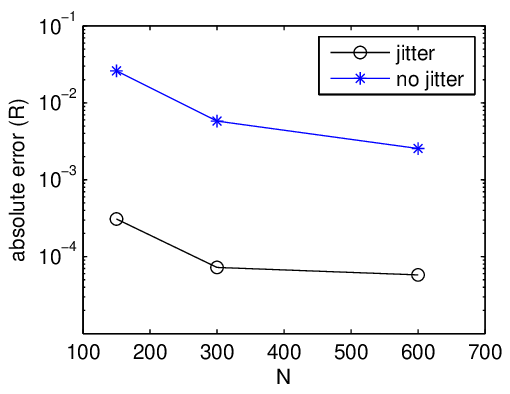}
        }
}
\centerline{
        \subfigure[Parameter $B$.]{
                \includegraphics[width=0.35\linewidth]{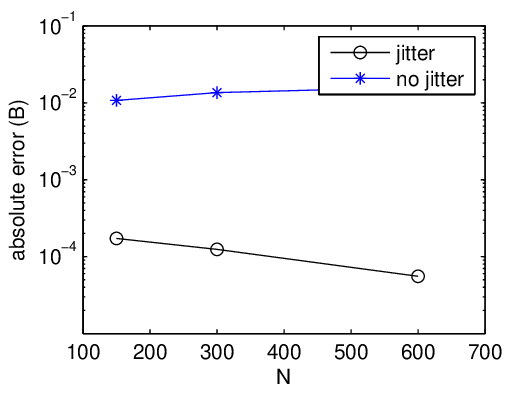}
        }
        \subfigure[Parameter $k_o$.]{
                \includegraphics[width=0.35\linewidth]{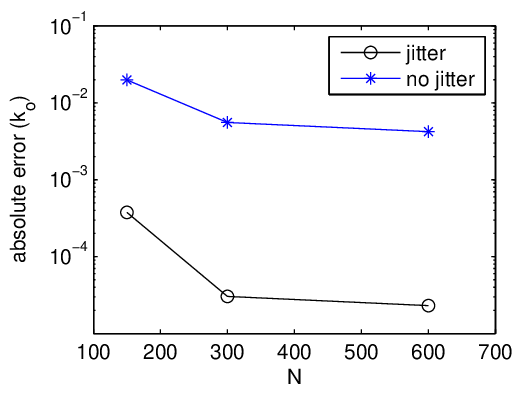}
        }
}
\caption{Empirical variance of the absolute parameter estimation errors over 20 independent simulation runs using Algorithm \ref{alRA} with $N=M=150$, $N=M=300$ and $N=M=600$ (the total number of particles is $N^2$). The errors are normalised w.r.t. the true parameter values, $S=10, R=28, B=\frac{8}{3}$ and $k_o=\frac{4}{5}$. The results obtained when jittering is suppressed in Algorithm \ref{alRA} (labeled as {\em no jitter}) are also shown for comparison. In each simulation, the Lorenz system was run for 24,000 discrete-time steps ($24$ continuous-time steps, for $\intstep=10^{-3}$), with observations collected every $40$ discrete steps.}
\label{fVarError_vs_N_t24}
\end{figure}

Figure \ref{fVarError_vs_N_t24} displays the empirical variance for the average errors of Figure \ref{fMeanError_vs_N_t24}, with and without jittering. It shows that the variability of the estimators is relatively large for small $t$ and it reduces considerably as a longer observation record is accumulated.

Finally, we have carried out a simple computer experiment to test the effect of a change-point in one of the parameters (the observation scale factor $k_o$). The simulation setup is the same as in the rest of this Section except that we extend the support of the parameter $k_o$ to be the interval $\left[\frac{1}{2},8\right]$, with uniform a priori probability distribution, and artificially introduce a change-point at continuous time instant 30, where $k_o$ changes its value from $0.8$ to $5$. This change-point is not described by the model, that represents $k_o$ as strictly constant. We have run Algorithm \ref{alRA} once, with $N=M=500$ particles, and observed the evolution over time of the posterior-mean estimators for $S$, $B$, $R$ and $k_o$.

\begin{figure}
\centerline{
        \subfigure[Parameter $S$.]{
                \includegraphics[width=0.35\linewidth]{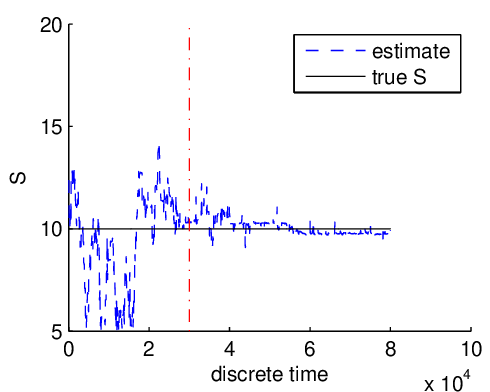}
        }
        \subfigure[Parameter $R$.]{
                \includegraphics[width=0.35\linewidth]{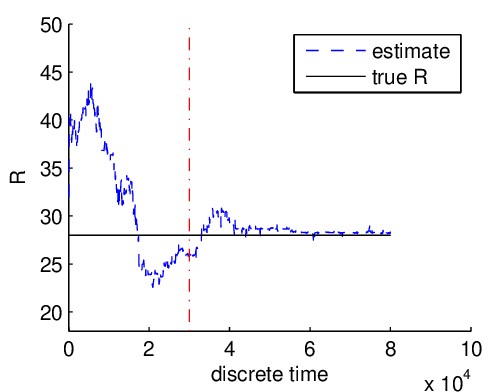}
        }
}
\centerline{
        \subfigure[Parameter $B$.]{
                \includegraphics[width=0.35\linewidth]{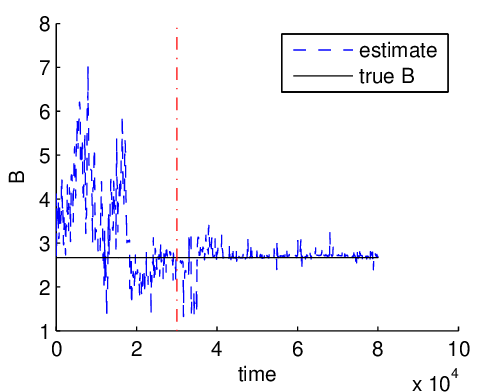}
        }
        \subfigure[Parameter $k_o$, with a change-point at time $3\times 10^4$.]{
                \includegraphics[width=0.35\linewidth]{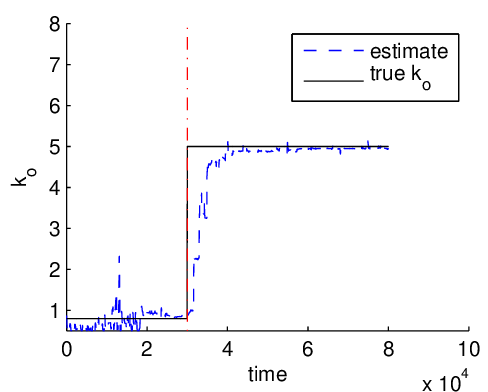}
        }
}
\caption{Evolution over time of the posterior-mean estimates of the parameters $S$, $B$, $R$ and $k_o$ for a single run of Algorithm \ref{alRA} with $N=M=500$. The actual parameter values of $S$, $R$, and $B$ are indicated with a horizontal solid line. The value of $k_o$ is also indicated, however it has a change-point at discrete time $3\times 10^4$ (from $0.8$ to 5). The change-point itself is marked by a vertical dashed line in the four plots. The algorithm is capable of tracking the change in $k_o$, however the adaptation of the estimator is limited by the variance of the jittering kernel and we observe a relatively long transition period of $\approx 10^4$ discrete time steps until the posterior mean settles around the new value.}
\label{fChange}
\end{figure}

Figure \ref{fChange} shows that the posterior-mean estimates fluctuate considerably for (relatively) small $t$, as we concluded from observing their empirical variance. The value of $k_o$ is changed at discrete time $3 \times 10^4$, which corresponds to continuous time 30 and a sequence of 750 observations. The change is instantaneous, yielding a step function for $k_o$ as plotted in Figure \ref{fChange}(d). Before the change-point, the random support of the posterior distribution of $k_o$ concentrates around the original value $k_o=0.8$. After the change-point, this support has to be adapted. However, the pace of this adaptation is limited by the variance of the jittering kernel and, hence, we observe a transition in the sequence of estimates that lasts for nearly $10^4$ time steps (10 continuous time units, 250 observations). Eventually, the posterior mean settles around the new value of $k_o$ in this simulation; however, further investigation is needed regarding the speed at which the random support of $\mu_t^{N,N}$ can be adapted and its interplay with estimation errors.

\section{Conclusions} \label{sConclusions}

We have introduced a recursive Monte Carlo scheme, consisting of two (nested) layers of particle filters, for the approximation and tracking of the posterior probability distribution of the unknown parameters of a state-space Markov system. Unlike existing SMC$^2$ and particle MCMC methods, the proposed algorithm is purely recursive and can be seen as a natural adaptation of the classic bootstrap filter to operate on the space of the static system-parameters.

The main theoretical contribution of the paper is the analysis of the errors in the approximation of integrals of bounded functions w.r.t. the posterior probability measure of the parameters. Using induction arguments, and placing only mild constraints on the state-space model and the parameters, we have proved that the $L_p$ norms of the approximation errors for the proposed algorithm vanish with rate proportional to $\frac{1}{\sqrt{N}} + \frac{1}{\sqrt{M}}$, where $N$ is the number of particles in the parameter space and $N \times M$ is the number of particles in the state space. This is achieved with a computational cost that grows only linearly with time. In comparison, the computational load of the SMC$^2$ method increases quadratically with time. The price to pay for this reduction in computational cost is that in the new scheme we need $N\rw\infty$ and $M\rw\infty$ in order to make the error converge towards 0, while the SMC$^2$ algorithm is consistent for fixed $M$, i.e., $N\rw\infty$ is sufficient for the errors to vanish, independently of $M$. As a consequence, if $K=NM$ is the total number of particles in the state space, then the optimal allocation for the proposed nested particle filter is $N=M=\sqrt{K}$ and the errors converge as $K^{-\frac{1}{4}}$ in $L_p$, while the SMC$^2$ scheme, with $M$ fixed, converges as $K^{-\frac{1}{2}}$.
 


The proposed algorithm can be combined with a SMC$^2$ scheme for practical convenience. For example, one may run a standard SMC$^2$ algorithm on the initial part of the observation sequence (possibly a few tens or a few hundreds of observations, depending on the problem and the available computational resources) to take advantage of its faster convergence rate and then switch to a recursive nested particle filter (Algorithm \ref{alRA}) when the computational cost of batch processing becomes too high. 

We also note that the continuity argument that leads to the derivation the the recursive nested particle filter, and the theoretical framework for the analysis of the resulting approximations, can be extended to other similar filtering algorithms. For example, it would be relatively straightforward to obtain a recursive version of the original IBIS algorithm of \citep{Chopin02}.

\section*{Acknowledgements} 

The work of the D. Crisan  has been partially supported by the EPSRC grant no EP/N023781/1. The work of J. M\'{\i}guez was partially supported by the Office of Naval Research Global (award no. N62909- 15-1-2011), {\em Ministerio de Econom\'{\i}a y Competitividad} of Spain (project TEC2015-69868-C2-1-R ADVENTURE) and {\em Ministerio de Educaci\'on, Cultura y Deporte} of Spain ({\em Programa Nacional de Movilidad de Recursos Humanos} PRX12/00690). 

Part of this work was carried out while J. M. was a visitor at the Department of Mathematics of Imperial College London, with partial support from an EPSRC Mathematics Platform grant. D. C. and J. M. would also like to acknowledge the support of the Isaac Newton Institute through the program ``Monte Carlo Inference for High-Dimensional Statistical Models'', as well as the constructive comments of an anonymous Reviewer, who helped improving the final version of this manuscript.

\appendix
\section{Proof of Lemma \ref{lm1StepUnifConv}} \label{ap1StepUnifConv}


We consider first the predictive measure 
$$
\xi_{t,\theta}^N(dx) = \frac{1}{N} \sum_{n=1}^N \delta_{\bar x_{t}^{(n)}}(dx)
$$
where $\bar x_{t}^{(n)}$, $n=1,...,N$, are the state particles drawn from the transition kernels $\tau_{t,\theta}^{x_{t-1}^{(n)}}(dx)  \dfn \tau_{t,\theta}(dx|x_{t-1}^{(n)})$ at the sampling step of the particle filter. Recall that $\xi_{t,\theta}=\tau_{t,\theta}\phi_{t-1,\theta}$ and consider the triangle inequality
\begin{eqnarray}
\left\|
	(f,\xi_{t,\theta}^N) - (f,\xi_{t,\theta})
\right\|_p &=& \left\|
	(f,\xi_{t,\theta}^N) - (f,\tau_{t,\theta}\phi_{t-1,\theta})
\right\|_p \nonumber \\
&\le& \left\|
	(f,\xi_{t,\theta}^N) - (f,\tau_{t,\theta}\phi_{t-1,\theta}^N)
\right\|_p \nonumber\\
&&+ \left\|
	(f,\tau_{t,\theta}\phi_{t-1,\theta}^N) - (f,\tau_{t,\theta}\phi_{t-1,\theta})
\right\|_p, \label{eqTriangle1}
\end{eqnarray}
where 
\begin{equation}
(f,\tau_{t,\theta}\phi_{t-1,\theta}^N) =  \frac{1}{N} \sum_{n=1}^N \int f(x) \tau_{t,\theta}(dx|x_{t-1}^{(n)}) 
= \frac{1}{N} \sum_{n=1}^N (f, \tau_{t,\theta}^{x_{t-1}^{(n)}}).
\label{eqIni}
\end{equation}
In the sequel we seek upper bounds for the $L_p$ norms in the right hand side of \eqref{eqTriangle1}.

Let us  introduce  the $\sigma$-algebra generated by the random paths $x_{0:t}^{(n)}$ and $\bar x_{1:t}^{(n)}$, $n=1,...,N$, denoted $\mF_t = \sigma\left( x_{0:t}^{(n)}, \bar x_{1:t}^{(n)}, \quad n=1,...,N \right)$. The conditional expectation of the integral $(f,\xi_{t,\theta}^N)$ given $\mF_{t-1}$ is
\begin{eqnarray}
E\left[
	(f,\xi_{t,\theta}^N) | \mF_{t-1}
\right] &=& \frac{1}{N} \sum_{n=1}^N E\left[
	f(\bar x_t^{(n)}) | \mF_{t-1}
\right] \nonumber \\
&=& \frac{1}{N} \sum_{n=1}^N (f, \tau_{t,\theta}^{x_{t-1}^{(n)}}) = (f,\tau_{t,\theta}\phi_{t-1,\theta}^N) \nonumber
\end{eqnarray}
and we note that the random variables $S_{t,\theta}^{(n)} =  f(\bar x_t^{(n)}) - (f,\tau_{t,\theta}^{x_{t-1}^{(n)}})$, $n = 1, ..., N$, are independent and zero-mean conditional on the $\sigma$-algebra $\mF_{t-1}$. For even $p$, the approximation error between $\xi_{t,\theta}^N$ and its (conditional) expected value $\tau_{t,\theta}\phi_{t-1,\theta}^N$ can then be written as
\begin{eqnarray}
E\left[
	\left(
		(f,\xi_{t,\theta}^N) - (f,\tau_t\phi_{t-1,\theta}^N) 
	\right)^p | \mF_{t-1}	
\right] &=& E\left[
	\left(
		\frac{1}{N} \sum_{n=1}^N S_{t,\theta}^{(n)}
	\right)^p | \mF_{t-1}
\right] \nonumber\\
&=& \frac{1}{N^p} \sum_{n_1=1}^N \cdots \sum_{n_p=1}^N E\left[
	S_{t,\theta}^{(n_1)}  \ldots  S_{t,\theta}^{(n_p)}
	| \mF_{t-1}
\right]. \nonumber\\\label{eqCombinations}
\end{eqnarray}
Since the random variables $S_{t,\theta}^{(n_i)}$ are conditionally independent and zero-mean, every term in the summation of \eqref{eqCombinations} involving a moment of order 1 vanishes. It is an exercise in combinatorics to show that the number of terms which {\em do not} contain any moment of order 1 is a polynomial function of $N$ with degree $\frac{p}{2}$, whose coefficients depend only on $p$. As a consequence, there exists a constant $\tilde c$ independent of $N$ such that the number of non-zero terms in \eqref{eqCombinations} is at most $\tilde c^p N^{\frac{p}{2}}$. Moreover, for each non-zero term we readily calculate the upper bound $
E\left[
	S_{t,\theta}^{(n_1)}  \ldots  S_{t,\theta}^{(n_p)}
	| \mF_{t-1}
\right] \le 2^p \| f \|_\infty^p
$. Therefore, for even $p$, we arrive at the inequality
\begin{equation}
E\left[
	\left(
		(f,\xi_{t,\theta}^N) - (f,\tau_t\phi_{t-1,\theta}^N) 
	\right)^p | \mF_{t-1}	
\right] \le \frac{\tilde c^p 2^p \| f \|_\infty^p}{N^\frac{p}{2}}
\label{eqBoundPredict-power-p}
\end{equation}
and taking unconditional expectations on both sides of \eqref{eqBoundPredict-power-p}, we readily find that, 
\begin{equation}
\| (f,\xi_{t,\theta}^N) - (f,\tau_t\phi_{t-1,\theta}^N \|_p \le \frac{c_1\|f\|_\infty}{\sqrt{N}},
\label{eqPredictionStep-1}
\end{equation}
where $c_1 = 2\tilde c$ is a constant independent of $N$ and $\theta$. The same inequality \eqref{eqPredictionStep-1} holds for any real $p$ because of the monotonicity of $L_p$ norms (an application of Jensen's inequality).

For the second term  in the right hand side of \eqref{eqTriangle1}, we note that $(f,\tau_{t,\theta}\phi_{t-1,\theta}) = (\bar f_\theta,\phi_{t-1,\theta})$, where $\bar f_\theta \in B(\mathbb{R}^{d_x})$ is a bounded\footnote{Trivially note that $\| \bar f_\theta \|_\infty \le \| f \|_{\infty}$, independently of $\theta$.} function defined as 
$$
\bar f_\theta(x) = \int f(x') \tau_{t,\theta}^x(dx') = (f,\tau_{t,\theta}^x)
$$ 
and, similarly, $(f,\tau_{t,\theta}\phi_{t-1,\theta}^N) = (\bar f_\theta,\phi_{t-1,\theta}^N)$. Therefore, assumption (\ref{eqInductionHypo}) yields the upper bound
\begin{eqnarray}
\left\|
	(f, \tau_{t,\theta} \phi_{t-1,\theta}^N) - (f,\tau_{t,\theta} \phi_{t-1,\theta})
\right\|_p &=& \left\|
	(\bar f_\theta, \phi_{t-1,\theta}^N) - (\bar f_\theta, \phi_{t-1,\theta})
\right\|_p Ê\nonumber \\
&\le& \frac{
	c_{t-1}\| f \|_\infty
}{
	\sqrt{N}
} +  \frac{
	\bar c_{t-1} \| f \|_\infty
}{
	\sqrt{M}
}, \label{eqConv_f_bar}
\end{eqnarray}
where the constants $c_{t-1}, \bar c_{t-1}$ are independent of $N$, $M$ and $\theta$. Substituting \eqref{eqPredictionStep-1} and \eqref{eqConv_f_bar} into (\ref{eqTriangle1}) yields
\begin{eqnarray}
\left\|
	(f,\xi_{t,\theta}^N) - (f,\xi_{t,\theta})
\right\|_p &\le& \frac{
	\tilde c_t\| f \|_\infty
}{
	\sqrt{N}
} +  \frac{
	\bar {\tilde c}_t\| f \|_\infty
}{
	\sqrt{M}
}, \label{eqBoundXi}
\end{eqnarray}
where $\tilde c_t = c_{t-1} + c_1$ and $\bar {\tilde c}_t = \bar c_{t-1}$ are finite constants independent of $N$, $M$ and $\theta$.

Next, we use inequality (\ref{eqBoundXi}) to calculate a bound on $\| (f,\phi_{t,\theta}^N) - (f,\phi_{t,\theta}) \|_p$. Let us first note that, after the computation of the weights, we obtain a random measure of the form 
$$
\bar \phi_{t,\theta}^N(dx) = \sum_{n=1}^N w_{t}^{(n)} \delta_{\bar x_{t}^{(n)}}(dx),
\quad \mbox{where}
\quad
w_{t}^{(n)} = \frac{
	g_{t,\theta}^{y_{t}}( \bar x_{t}^{(n)} )
}{
	\sum_{k=1}^N g_{t,\theta}^{y_{t}}(\bar x_{t}^{(n)})
}. 
$$
As a consequence, integrals w.r.t. the measure $\bar \phi_{t,\theta}^N$ can be written in terms of $g_{t,\theta}^{y_{t}}$ and $\xi_{t,\theta}^N$, namely
\begin{equation}
(f,\bar \phi_{t,\theta}^N) = \frac{
	(fg_{t,\theta}^{y_{t}},\xi_{t,\theta}^N)
}{
	(g_{t,\theta}^{y_{t}},\xi_{t,\theta}^N)
}.
\label{eqLa1era}
\end{equation}
This is natural, though, since from the Bayes theorem we readily derive the same relationship between $\phi_{t,\theta}$ and $\xi_{t,\theta}$,
\begin{equation}
(f,\phi_{t,\theta}) = \frac{
	(fg_{t,\theta}^{y_{t}},\xi_{t,\theta})
}{
	(g_{t,\theta}^{y_{t}},\xi_{t,\theta})
}.
\label{eqLa2a}
\end{equation}
Given \eqref{eqLa1era} and \eqref{eqLa2a}, we can readily apply the inequality \eqref{eqPreliminaries} to obtain 
\begin{eqnarray}
\left|
	(f,\bar \phi_{t,\theta}^N) - (f,\phi_{t,\theta}) 
\right| &\le& \frac{
	1
}{
	(g_{t,\theta}^{y_{t,\theta}},\xi_{t,\theta})
} \left(
	\| f \|_\infty \left|
		(g_{t,\theta}^{y_{t}},\xi_{t,\theta}) - (g_{t,\theta},\xi_{t,\theta}^N)
	\right| 
\right. \nonumber\\
&&\left. 
	+  \left|
		(fg_{t,\theta}^{y_{t}},\xi_{t,\theta}) - (fg_{t,\theta},\xi_{t,\theta}^N)
	\right|
\right),
\label{eqSubtraction-2}
\end{eqnarray}
where $u_t(\theta) = (g_{t,\theta}^{y_{t,\theta}},\xi_{t,\theta}) > 0$ by assumption. From \eqref{eqSubtraction-2} and Minkowski's inequality,
\begin{eqnarray}
\left\|
	(f,\bar \phi_{t,\theta}^N) - (f,\phi_{t,\theta}) 
\right\|_p &\le&  \frac{
	1
}{
	(g_{t,\theta}^{y_{t,\theta}},\xi_{t,\theta})
} \times \left(
	\| f \|_\infty \left\|
		(g_{t,\theta}^{y_{t}},\xi_{t,\theta}) - (g_{t,\theta}^{y_t},\xi_{t,\theta}^N)
	\right\|_p 
\right.\nonumber \\
&& \left.
	+ \left\|
		(fg_{t,\theta}^{y_{t}},\xi_{t,\theta}) - (fg_{t,\theta},\xi_{t,\theta}^N)
	\right\|_p
\right) 
\label{eqSubtraction-3}
\end{eqnarray}
and, since $\| g_{t,\theta}^{y_t}\|_\infty \le \| g_t^{y_t} \|_\infty < \infty$ by assumption (in particular, $\| g_t^{y_t} \|_\infty$ is independent of $\theta$), the inequalities \eqref{eqBoundXi} and \eqref{eqSubtraction-3} together yield
\begin{equation}
\left\|
	(f,\bar \phi_{t,\theta}^N) - (f,\phi_{t,\theta}) 
\right\|_p \le  \frac{
	2\| f \|_\infty \| g_t^{y_t} \|_\infty \tilde c_t
}{
	(g_{t,\theta}^{y_{t,\theta}},\xi_{t,\theta})
} \times \frac{
	1
}{
	\sqrt{N}
} + \frac{
	2\| f \|_\infty \| g_t^{y_t} \|_\infty \bar {\tilde c}_t
}{
	(g_{t,\theta}^{y_{t,\theta}},\xi_{t,\theta})
} \times \frac{
	1
}{
	\sqrt{M}
}, \label{eqSubtraction-4}
\end{equation}
where the finite constants $\tilde c_t$ and $\bar {\tilde c}_t = \bar c_{t-1}$ are independent of $N$, $M$ and $\theta$. Indeed, the only factor that depends on $\theta$ in the right-hand side of \eqref{eqSubtraction-4} is the integral $u_t(\theta) = (g_{t,\theta}^{y_{t,\theta}},\xi_{t,\theta})$. However, we have assumed that 
\begin{equation}
u_{t,\inf} = \inf_{\theta \in D_\theta} u_t(\theta) > 0,
\end{equation}
hence the inequality \eqref{eqSubtraction-4} leads to
\begin{equation}
\left\|
	(f,\bar \phi_{t,\theta}^N) - (f,\phi_{t,\theta}) 
\right\|_p \le  \frac{
	c_{2,t}\| f \|_\infty
}{
	\sqrt{N}
} + \frac{
	\bar c_{2,t} \| f \|_\infty
}{
	\sqrt{M}
}\label{eqBoundForWeighted}
\end{equation}
where 
\begin{equation}
c_{2,t} =  \frac{
	2 \| g_t^{y_t} \|_\infty \tilde c_t
}{
	u_{t,\inf}
} < \infty 
\quad \mbox{and} \quad 
\bar c_{2,t} =  \frac{
	2 \| g_t^{y_t} \|_\infty \bar c_{t-1}
}{
	u_{t,\inf}
} < \infty
\label{eqPolaCte}
\end{equation}
are constants independent of $N$, $M$ and $\theta$.

Finally, we only need to verify the resampling step, i.e., that the $L_p$ norm $\| (f,\phi_{t,\theta}^N) - (f,\bar \phi_{t,\theta}^N) \|_p$ is bounded as well. Let $\bar \mF_{t} = \sigma\left( x_{0:t-1}^{(n)}, \bar x_{1:t}^{(n)}; n=1,\ldots,N \right)$ be the $\sigma$-algebra generated by the random sequences $x_{0:t-1}^{(n)}$ and $\bar x_{1:t}^{(n)}$, $n=1,...,N$. It is straightforward to check that, for every $n=1, ..., N$,
\begin{equation}
E\left[
	f(x_t^{(n)}) | \bar \mF_t
\right] = (f, \bar \phi_{t,\theta}^N),
\label{eqPpioResampling}
\end{equation}
hence the random variables $\bar S_{t,\theta}^{(n)} = f(x_t^{(n)}) - (f, \bar \phi_{t,\theta}^N)$ are independent and zero-mean conditional on the $\sigma$-algebra $\bar \mF_t$. Therefore, the same combinatorial argument that led to Eq. \eqref{eqPredictionStep-1} now yields
\begin{equation}
\left\|
	(f,\phi_{t,\theta}^N) - (f,\bar \phi_{t,\theta}^N)
\right\|_p \le \frac{
	c_3 \| f \|_\infty
}{
	\sqrt{N}
} \label{eqBoundForResampling}
\end{equation}
where the constant $c_3$ is independent of both $N$ and $\theta$ (it does not depend on the distribution of the error variables $\bar S_{t,\theta}^{(n)}$). Since
\begin{equation}
\| (f,\phi_{t,\theta}^N) - (f,\phi_{t,\theta}) \|_p \le \| (f,\phi_{t,\theta}^N) - (f,\bar \phi_{t,\theta}^N) \|_p + \| (f,\bar \phi_{t,\theta}^N) - (f,\phi_{t,\theta}) \|_p,
\label{eqYetAnotherTriangle}
\end{equation}
substituting Eqs. \eqref{eqBoundForResampling} and \eqref{eqBoundForWeighted} into the inequality \eqref{eqYetAnotherTriangle} yields
\begin{equation}
\| (f,\phi_{t,\theta}^N) - (f,\phi_{t,\theta}) \|_p \le \frac{
	c_t\| f \|_\infty
}{
	\sqrt{N}
} + \frac{
	\bar c_t\| f \|_\infty
}{
	\sqrt{M}
}, \label{eqFinalResampling}
\end{equation}
where $c_t = c_3 + c_{2,t}$ and $\bar c_t = \bar c_{2,t}$ are finite constants independent of both $N$, $M$ and $\theta$. 

To complete the proof, simply note that $\bar c_{t-1}=0$ implies $\bar c_t = \bar c_{2,t} = 0$ (see \eqref{eqPolaCte}). 
$\QED$

\section{Proof of Lemma \ref{lmUnifConvPF}} \label{apProof_of_Lemma_2}

We proceed by induction in $t$. For $t=0$, the measure $\phi_{0,\theta}^N(dx) = \frac{1}{N} \sum_{n=1}^N \delta_{x_0^{(n)}}(dx)$ is constructed from an i.i.d. sample of size $N$ from the prior distribution $\phi_{0,\theta} \equiv \tau_0$. Then, it is straightforward to prove that
$$
\| (f,\phi_{0,\theta}^N) - (f,\phi_{0,\theta}) \|_p \le \frac{
	c_0\| f \|_\infty
}{
	\sqrt{N}
},
$$
where $c_0<\infty$ is independent of $N$. Note that, since $\phi_{0,\theta} \equiv \tau_{0}$ is actually independent of $\theta$, the constant $c_0$ is independent of $\theta$ as well. 

For the induction step, we assume that
\begin{equation}
\| (f,\phi_{t-1,\theta}^N) - (f,\phi_{t-1,\theta}) \|_p \le \frac{
	c_{t-1}\| f \|_\infty
}{
	\sqrt{N}
} \label{eqInductionHypo-2}
\end{equation}
holds true for some constant $c_{t-1}<\infty$ independent of $N$ and $\theta$. Given \eqref{eqInductionHypo-2}, Lemma \ref{lm1StepUnifConv} yields 
$$
\| (f,\xi_{t,\theta}^N) - (f,\xi_{t,\theta}) \|_p \le \frac{
	\tilde c_t \| f \|_\infty
}{
	\sqrt{N}
} \quad \mbox{and} \quad
\| (f,\phi_{t,\theta}^N) - (f,\phi_{t,\theta}) \|_p \le \frac{
	c_t \| f \|_\infty
}{
	\sqrt{N}
}
$$
at time $t$, where $\tilde c_t$ and $c_t$ are finite constants independent of $N$ and $\theta$. 
$\QED$

\section{A family of jittering kernels} \label{apPequeLema}

\begin{Proposicion} \label{propPeque}
Assume that $h \in B(D_\theta)$ is Lipschitz, with constant $c_L\|h\|_\infty < \infty$, and consider the class of kernels $\kappa_N^{\theta'} = (1-\epsilon_N)\delta_{\theta'} + \epsilon_N\bar \kappa_N^{\theta'}$, where $0\le\epsilon_N\le 1$ and $\bar \kappa_N^{\theta'} \in \mP(D_\theta)$. For any $p \ge 1$, if the kernel $\kappa_N^{\theta'}$ is selected in such a way that
\begin{equation}
\sigma_{\kappa,N}^2 = \sup_{\theta'\in D_\theta} \int \| \theta - \theta' \|^2 \bar \kappa_N^{\theta'}(d\theta) \le \frac{
	\breve c
}{
	\epsilon_N^\frac{p+2}{p} N^\frac{p+2}{2}
}
\label{eqIneqSigmaKappa_p}
\end{equation}
is satisfied for some constant $\breve c < \infty$ independent of $N$, then the inequality
\begin{equation}
\sup_{\theta' \in D_\theta} \int | h(\theta) - h(\theta') |^p \kappa_N^{\theta'}(d\theta) \le \frac{
	c_\kappa^p\|h\|_\infty^p
}{
	N^\frac{p}{2}
}
\nonumber
\end{equation}
holds for a constant $c_\kappa^p= c_L^p \left(
	1 + \breve c \sup_{\theta_1,\theta_2 \in D_\theta} \| \theta_1 - \theta_2 \|^p
\right) < \infty$ independent of $N$.
\end{Proposicion}

\noindent {\bf Proof.} Since $\kappa_N^{\theta'} = (1-\epsilon_N)\delta_{\theta'} + \epsilon_N\bar \kappa_N^{\theta'}$ and $h$ is Lipschitz with constant $c_L\|h\|_\infty<\infty$, we readily obtain
\begin{equation}
\int |h(\theta)-h(\theta')|^p\kappa_N^{\theta'}(d\theta) \le \epsilon_N c_L^p\|h\|_\infty^p \int \| \theta - \theta' \|^p \bar \kappa_N^{\theta'}(d\theta).
\label{eqPeque1}
\end{equation}
Let 
\begin{equation}
\beta_N=\frac{1}{\epsilon_N^\frac{1}{p} \sqrt{N}}.
\label{eqBetaN}
\end{equation} 
We can rewrite \eqref{eqPeque1} as
\begin{eqnarray}
\int |h(\theta)-h(\theta')|^p \kappa_N^{\theta'}(d\theta) &\le& \epsilon_N c_L^p \|h\|_\infty^p \left[
	\int I_{\theta \in D_\theta: \| \theta - \theta' \| < \beta_N}(\theta) \| \theta - \theta' \|^p \bar \kappa_N^{\theta'}(d\theta)
\right. \nonumber\\
&& \left.
	+ \int I_{ \theta \in D_\theta: \| \theta - \theta' \| \ge \beta_N }(\theta) \| \theta - \theta' \|^p \bar \kappa_N^{\theta'}(d\theta)
\right] \nonumber\\
&\le& \epsilon_N c_L^p \|h\|_\infty^p \left[
	\beta_N^p + \hat C^p \int I_{ \theta \in D_\theta: \| \theta - \theta' \| \ge \beta_N }(\theta) \bar \kappa_N^{\theta'}(d\theta)
\right], \nonumber\\
\label{eqPeque2}
\end{eqnarray}
where $\hat C^p = \sup_{\theta_1, \theta_2 \in D_\theta} \| \theta_1 - \theta_2 \|^p < \infty$, since $D_\theta$ is compact. Using Chebyshev's inequality on the right hand side of \eqref{eqPeque2} yields
\begin{equation}
\int |h(\theta)-h(\theta')|^p \kappa_N^{\theta'}(d\theta) \le \epsilon_N c_L^p \|h\|_\infty^p \left(
	\beta_N^p + \hat C^p \frac{
		\sigma_{\kappa,N}^2
	}{
		\beta_N^2
	}
\right)
\label{eqPeque3}
\end{equation}
and substituting \eqref{eqIneqSigmaKappa_p} and \eqref{eqBetaN} into \eqref{eqPeque3} we arrive at
\begin{equation}
\int |h(\theta)-h(\theta')|^p \kappa_N^{\theta'}(d\theta) \le \frac{
	c_L^p \|h\|_\infty^p \left(
		1 + \breve c \hat C^p
	\right)
}{
	N^\frac{p}{2}
},\nonumber
\end{equation}
where all the constants are independent of $\theta'$ and $N$. 
$\QED$

\begin{Corolario}
Consider the same class of kernels $\kappa_N^{\theta'} = (1-\epsilon_N)\delta_{\theta'} + \epsilon_N\bar \kappa_N^{\theta'}$, where $0\le\epsilon_N\le 1$ and $\bar \kappa_N^{\theta'} \in \mP(D_\theta)$. For any $p \ge 1$, if \eqref{eqIneqSigmaKappa_p} holds for some $\breve c < \infty$ independent of $N$ then 
\begin{equation}
\sup_{\theta' \in D_\theta} \int \| \theta - \theta' \|^p \kappa_N^{\theta'}(d\theta) \le \frac{
	c_\kappa^p
}{
	N^\frac{p}{2}
}
\nonumber
\end{equation}
where $c_\kappa^p = 1 + \breve c \sup_{\theta_1,\theta_2 \in D_\theta} \| \theta_1 - \theta_2 \|^p < \infty$ is constant and independent of $N$.
\end{Corolario}

\noindent {\bf Proof.} Simply note that 
$$
\int \| \theta - \theta' \|^p \kappa_N^{\theta'}(d\theta) \le
\epsilon_N \int \| \theta - \theta' \|^p \bar \kappa_N^{\theta'}(d\theta)
$$
and then follow the same argument as in the proof of Proposition \ref{propPeque}.
$\QED$

\bibliographystyle{imsart-nameyear}
\bibliography{bibliografia}

\end{document}